\documentclass[aoas,MSNbibl,nameyear,rotating,seceqn,dvips]{arximspdf}
\usepackage{dcolumn}
\usepackage{graphicx}

\doi{10.1214/12-AOAS621} 
\volume{7}
\issue{2}
\pubyear{2013}
\firstpage{1217}
\lastpage{1243}

\makeatletter
\newcolumntype{d}[1]{D{.}{.}{#1}}
\def\R{\mathbb{R}}
\def\ln{\operatorname{ln}}

\newproclaim{remark}{Remark}
\newproclaim{definition}{Definition}
\makeatother

\begin{document}
\begin{frontmatter}

\title{A reference-invariant health disparity index based on R\'{e}nyi
divergence}
\runtitle{Symmetrized R\'{e}nyi index}

\begin{aug}
\author{\fnms{Makram} \snm{Talih}\corref{}\thanksref{t1}\ead[label=e1]{mtalih@cdc.gov}}
\thankstext{t1}{The findings and conclusions in this paper are those of
the author and do not
necessarily represent the views of the CDC/National Center for Health
Statistics. This
work was performed while the author was on sabbatical leave from the
City University of
New York School of Public Health at Hunter College, where he was
Associate Professor of Epidemiology and Biostatistics.}
\runauthor{M. Talih}
\affiliation{National Center for Health Statistics}
\address{Statistician (Health)---Senior Service Fellow\\
Centers for Disease Control and Prevention\\
National Center for Health Statistics\\
Office of Analysis and Epidemiology\\
Health Promotion Statistics Branch\\
3311 Toledo Road, Room 6317\\
Hyattsville, Maryland, 20782\\
USA\\
\printead{e1}}
\end{aug}

\received{\smonth{9} \syear{2012}}
\revised{\smonth{12} \syear{2012}}

%
\begin{abstract}
One of four overarching goals of Healthy People 2020 (HP2020) is to
achieve health equity, eliminate disparities, and improve the health of
all groups. In health disparity indices (HDIs) such as the mean log
deviation (MLD) and Theil index (TI), disparities are relative to the
population average, whereas in the index of disparity (IDisp) the
reference is the group with the least adverse health outcome. Although
the latter may be preferable, identification of a reference group can
be affected by statistical reliability. To address this issue, we
propose a new HDI, the R\'{e}nyi index (RI), which is
reference-invariant. When standardized, the RI extends the Atkinson
index, where a disparity aversion parameter can incorporate societal
values associated with health equity. In addition, both the MLD and TI
are limiting cases of the RI. Also, a symmetrized R\'{e}nyi index (SRI)
can be constructed, resulting in a symmetric measure in the two
distributions whose relative entropy is being evaluated. We discuss
alternative symmetric and reference-invariant HDIs derived from the
generalized entropy (GE) class and the Bregman divergence, and argue
that the SRI is more robust than its GE-based counterpart to small
changes in the distribution of the adverse health outcome. We evaluate
the design-based standard errors and bootstrapped sampling
distributions for the SRI, and illustrate the proposed methodology
using data from the National Health and Nutrition Examination Survey
(NHANES) on the 2001--04 prevalence of moderate or severe periodontitis
among adults aged 45--74, which track Oral Health objective OH-5 in
HP2020. Such data, which use a binary individual-level outcome
variable, are typical of HP2020 data.
\end{abstract}

%

\end{frontmatter}

\section{Background and introduction} \label{intro}

The measurement, tracking, and elimination of health disparities are
central to the U.S. Healthy People initiative; see \citet
{greenfielding2011}. One of two overarching goals of Healthy People
2010 (HP2010) was to ``eliminate health disparities'' [U.S. Department of Health and Human
Services (\citeyear{hp20102000,mcr2006}); National Center for Health
Statistics (\citeyear{fr2011})], and one of four overarching goals of
Healthy People 2020 (HP2020) is to ``achieve health equity, eliminate
disparities, and improve the health of all groups'' (\url{http://healthypeople.gov}).
There are several concepts and definitions
associated with the terms \textit{health disparities} and \textit{health
equity,} which are reviewed in \citet{braveman2006}. In this paper, we
do not discuss how to measure or assess health equity. Instead, we
restrict our attention to the measurement of health disparities,
although, as seen below, measures of health disparities are inevitably
tied to normative or societal values associated with health equity. Our
working definition of \textit{health disparity} is that of \citet
{keppelpearcyklein2004}, who state that ``in the context of public
health, a \textit{disparity} is the quantity that separates a group from a
reference point on a particular measure of health that is expressed in
terms of a rate, proportion, mean, or some other quantitative measure.''

When there are three or more population groups, for example, population
breakdown by race and ethnicity, education, or income, the differences
among those groups in the magnitude of their disparities relative to
the reference point can be summarized using a between-group index. Such
between-group health disparity indices (HDIs) have been reviewed in
\citet{wagstaffetal1991}, \citet{mackenbachkunst1997}, and \citet
{pearcykeppel2002}. Their characteristics and limitations have been
investigated in \citet{keppelpearcyklein2004}, \citet
{keppelpamuklynchetal2005}, \citet{levyetal2006}, and \citeauthor{harperetal2008} (\citeyear{harperetal2008,harperetal2010}).

For a population that is partitioned into $m$ mutually exclusive groups
of sizes $n_1, n_2, \ldots, n_m$, with $n = \sum_{j=1}^m n_j$, we
study the distribution of a particular adverse health outcome, which,
at the individual level, is given by $y_{ij}$, say, for individual~$i$
in group $j$. Specifically, our goal is to compare the aggregate health
outcomes $y_{\cdot j} = \sum_{i=1}^{n_j} y_{ij}$, $j=1, 2, \ldots, m$,
across groups. When the variable $y_{ij}$ is a binary variable,
indicating presence or absence of the adverse health outcome for
individual $i$, the aggregate $y_{\cdot j}$ is simply the frequency
count of the number of individuals in group $j$ with the adverse health outcome.

We look upon (between-group) HDIs as measures of generalized relative
entropy (or divergence) between two nonnegative mass functions $p =
(p_1, p_2, \ldots, p_m)$ and $q = (q_1, q_2, \ldots,  q_m).$ In the
analysis of health disparities, the quantities $p_j$ can be weights
that the analyst assigns to each population group $j.$ The groups are
said to be ``equally-weighted'' if they are assigned equal weights (e.g.,
$1/m$) and ``population-weighted'' if they are assigned weights that are
proportional to their size (e.g., $n_j/n$). On the other hand, the
$q_j$ can quantify the disease burden in group $j.$ Various HDIs differ
in the specification of the $q_j.$ Entropy-based HDIs commonly specify
$q_j$ as a function of the ratio between the group average ($\bar
{y}_{\cdot j}$) and a fixed reference for measuring disparities. The
reference can be the population average ($\bar{y}_{\cdot\cdot}$), the
least adverse health outcome ($\min_{1\leq k \leq m} \bar{y}_{\cdot
k}$), a Healthy People target, or any other reference.

In this paper, we introduce a new class of HDIs, the R\'{e}nyi index
(RI), which is based on a generalized R\'{e}nyi (or alpha--gamma)
divergence. Generalized R\'{e}nyi divergence was considered by \citet
{fujisawaeguchi2008} in the context of robust parameter estimation in
the presence of outliers and is reviewed in \citet{cichockiamari2010}.
The RI is a class of HDIs that are invariant to the choice of the
reference used for evaluating disparities. This invariance
property---also known as ``strong scale-invariance''---is relevant to
Healthy People, as well as to other initiatives that monitor health
disparities, because the identification of a reference group can be
affected by statistical reliability; see \citet{fr2011}.
Reference-invariance is not unique to the RI. As discussed in Section
\ref{alternatives}, the well-known generalized entropy (GE) class, for
one, can be modified for strong scale-invariance. Nonetheless, the
robustness of the RI makes it less sensitive than its GE-based
counterpart to small changes in the distribution of the adverse health outcome.

Looking at HDIs as measures of generalized relative entropy (or
divergence) between two nonnegative mass functions $p$ and $q$---not
necessarily probabilities---provides a common mathematical framework
within which various HDIs can be compared. In particular, this unified
framework enables a sensitivity analysis for the effect of changing the
reference used for evaluating disparities (e.g., average versus best
group rate) as well as the effect of modifying the weighting
distribution~$p_j$ (equally-weighted versus population-weighted), which
are issues of concern; see \citet{harperetal2010}.

The RI is a class of HDIs, $\{\operatorname{RI}_{\alpha}\dvtx \alpha\in\R\}.$ When
the parameter $\alpha> 0$ increases, the rescaled index $\alpha\operatorname
{RI}_{\alpha}$ is nondecreasing; therefore, $\alpha$ can be interpreted
as a disparity aversion parameter in a manner akin to the Atkinson
index [\citeauthor{atkinson1970} (\citeyear{atkinson1970})]. Indeed,
for $\alpha> 0$, the Atkinson index simply is obtained via the
standardizing exponential transformation $1-e^{-\alpha\operatorname{RI}_{\alpha}}$.
The disparity aversion parameter can reflect a range of societal
values attached to inequality. In \citet{levyetal2006}, the Atkinson
index is shown to fulfill some of the core axioms of health benefits
analysis, for example, Pigou--Dalton transfer principle and subgroup
decomposability. The authors also argue that, unlike some indices in
the GE class, the Atkinson index avoids a value judgment about the
relative importance of transfers at different percentiles of the
distribution of the adverse health outcome.

In this paper, we illustrate the proposed methodology using data from
the National Health and Nutrition Examination Survey (NHANES) on the
2001--04 prevalence of moderate or severe periodontitis among U.S.
adults aged 45--74. These binary individual-level data track Oral
Health objective OH-5 in HP2020. NHANES is the data source for
approximately 1 in 7 population-based objectives in HP2020. Close to
half of the (approximately) 1200 objectives in HP2020 are
population-based, and most, though not all, such objectives track a
proportion or a rate where the underlying individual-level variable has
a binary outcome. See \url{http://healthypeople.gov}. The supplement to
this article in \citet{talih2012suppDT} provides further illustration
of the proposed methodology with continuous individual-level data on
total blood cholesterol levels among adults aged 20 and over, from
NHANES 2005--08. These data track Heart Disease and Stroke objective
HDS-8 in HP2020.

\subsection{Common practice}\label{common-practice}

The most commonly used between-group HDIs are weighted sums of the form
$\sum_{j=1}^m p_j f(r_j)$, where $r_j = q_j/p_j$, for some function~$f(r);$ see \citet{firebaugh1999}.

\subsubsection*{Population average as reference}

When the distributions $p$ and $q$ are given by
$p_j = n_j/n$, $q_j = p_jr_j$, and $r_j$ is the ratio of the average
$\bar{y}_{\cdot j}$ of the adverse health outcome in group $j$ relative
to the population average $\bar{y}_{\cdot\cdot}$,
%
\begin{equation}
\label{intro-rj} r_j = \bar{y}_{\cdot j}/\bar{y}_{\cdot\cdot},
\end{equation}
the resulting class of HDIs, with $f(r) = f_{\alpha}(r):= (1-r^{1-\alpha})/[\alpha(1-\alpha)]$, is the generalized entropy (GE) class, which
extends the mean log deviation [MLD; $f_1(r) = -\ln r$] and the Theil
index [TI; $f_0(r) = r\ln r$]; see \citeauthor{worldbank2009}
[(\citeyear{worldbank2009}); Chapter 6].

That the GE HDIs are nonnegative, that equal zero only when $r_j=1$
for all~$j$, follows from the convexity of the function $f_{\alpha}(r)$
specified above and the fact that the $p_j$ and $q_j$ sum to one---GE
is a class of Csisz\'{a}r $f$-divergences; see \citet{alisilvey1966}.
However, the requirement that the distributions $p$ and $q$ be
probability mass functions can be restrictive.

\subsubsection*{Least adverse health outcome as reference}

Keppel
et~al. (\citeyear{keppelpamuklynchetal2005}) recommend measuring disparities
relative to the group with the least adverse health outcome. Instead of
(\ref{intro-rj}), this would result in $r_j$'s of the form
%
\begin{equation}
\label{intro-rj2} r_j = \frac{\bar{y}_{\cdot j}}{\min_{1\leq k \leq m}\bar{y}_{\cdot k}}.
\end{equation}
Clearly, with the $r_j$ as in (\ref{intro-rj2}) and the $p_j=1/m$ or
$n_j/n$, the $q_j = p_j r_j$ no longer define a probability mass function.

\subsubsection*{The health inequality paradox}

There are two essentially distinct approaches to evaluating health
disparities overall, each of which makes an explicit value judgment
regarding the trade-off between an individual's burden of disease and a
group's burden of disease. Used in the GE class, which includes the MLD\vadjust{\goodbreak}
and TI, the population-weighted distribution $p_j = n_j/n$ is
consistent with individuals in the population being
equally-weighted---with weights $1/n$---regardless of their group
membership. In contrast, the equally-weighted distribution $p_j=1/m$,
which is used in the index of disparity (IDisp) of \citet
{keppelpamuklynchetal2005}, results in more weight being given to
individuals in smaller population groups than in larger ones.
[\citeauthor{keppelpamuklynchetal2005} use weights $1/(m-1)$
instead of $1/m$, since there are only $m-1$ comparisons relative to
the group with the least adverse health outcome.]

Because of this trade-off between the individual's burden of disease
and the group's burden of disease, potential impact of public health
interventions is modified by the specific measure of health disparities
used. When all groups are equally-weighted, an intervention that
targets a relatively small group with a relatively large burden of
disease can prove very effective in reducing overall disparity. On the
other hand, when groups are population-weighted, that same intervention
will not have as much impact on reducing overall disparity, and other
interventions might be desired; see \citet{harperetal2010}, \citet
{frohlichpotvin2008}, and \citet{rose1985}. The aforementioned
trade-off between individual's health and population's health is,
perhaps, what differentiates most strikingly analyses of health
disparities from studies of wealth inequalities---the latter having
provided the impetus for the development of the GE and related families
of inequality indices. When possible, methods used for comparing health
outcomes should similarly be differentiated from those used for
comparing income distributions.

\subsection{Organization of the paper}

The paper is organized as follows.

In Section \ref{G} we introduce the generalized R\'{e}nyi divergence as
the basis for developing the RI. A critical property of the generalized
R\'{e}nyi divergence is its invariance to scaling of either of the two
distributions whose divergence is being evaluated; see Section \ref
{G1}. Thus, when monitoring health disparities, the RI remains the
same, regardless of whether we use: the population average as the
denominator for the relative disparities $r_j$, as in (\ref{intro-rj});
the group with the least adverse health outcome as the denominator, as
in (\ref{intro-rj2}); or use a Healthy People or some other target as
the denominator.

The RI extends the MLD and TI. The MLD is mostly influenced by groups
with large population shares $p_j = n_j/n$, whereas the TI is mostly
influenced by groups where the adverse health outcome is more frequent
or severe ($q_j = y_{\cdot j}/y_{\cdot\cdot}$); see Section \ref{G0}.
In Section \ref{G3} we show that the RI can be symmetrized, which
yields a symmetric measure in the two distributions whose relative
entropy is being evaluated. Thus, when $p_j = n_j/n$, $q_j=p_jr_j$, and
the $r_j$ are as in (\ref{intro-rj}), the resulting symmetrized R\'
{e}nyi index (SRI) generalizes the symmetrized Theil index (STI) of
\citet{borrelltalih2011}.

In Section \ref{G2} we show that, for $\alpha> 0$, the Atkinson index
simply is obtained via the standardizing exponential transformation
$1-e^{-\alpha\operatorname{RI}_{\alpha}}$. Hence, the parameter $\alpha>0$ is a
disparity aversion parameter for $\alpha\operatorname{RI}_{\alpha}$ and it can
reflect a range of societal values attached to inequality.\vadjust{\goodbreak}

Because of scale invariance, the RI and SRI only depend on the relative
disparities $r_j$ in (\ref{intro-rj}) or (\ref{intro-rj2}) through the
numerator $\bar{y}_{\cdot j}.$ Thus, in Section~\ref{SRIb}, we express
the between-group RI and SRI as functions of the group sizes $n_j$ and
means $\bar{y}_{\cdot j}$, both when groups are population-weighted
($p_j = n_j/n$) and when groups are equally-weighted ($p_j = 1/m$).

In Section \ref{alternatives} we discuss two potential alternatives to
the RI based on the GE (Section \ref{ALPHA}) and Bregman class (Section
\ref{BETA}). In Section \ref{sim-studies} we compare the
(reference-invariant) SRI with a symmetrized reference-invariant GE
under simple hypothetical scenarios and argue that the SRI is less
sensitive to small changes in the distribution of the adverse health outcome.

In Section \ref{stderrors} we proceed as in \citet{borrelltalih2011}
and \citet{biewenjenkins2006} to derive design-based standard errors
for the (between-group) RI and SRI using Taylor series linearization.
To validate our derivation, we implement in the supplemental R code the
balanced repeated replication and bootstrap methods, introduced by \citet
{mccarthy1969} and \citet{raowu1988}, respectively; see \citet
{talih2012suppRR}. Rescaled bootstrap enables the design-based
estimation of the sampling distribution of the RI and SRI. Further, we
examine the effect of the weighting distribution $p$, comparing the
population-weighted ($p_j=n_j/n$) to the equally-weighted case ($p_j = 1/m$).

In Section \ref{case-studies} we illustrate the proposed methodology
using periodontal disease data from NHANES. Section \ref{DISC} concludes.

The technical appendix includes a detailed discussion of the
decomposability of the RI and SRI; see \citet{talih2012suppTA}.
Decomposability is the separation of the total or aggregate HDI into
between- and within-group components; see \citet{bourguignon1979}. Just
like for the GE class of HDIs, decomposability allows for multiple
predictors of individual-level disparities to be considered in
succession, as in multi-way analysis of variance. We examine the
decomposition of the total RI and SRI when groups are
population-weighted (e.g., $p_j = n_j/n$)---which, as mentioned
earlier, is consistent with individuals being equally-weighted---as
well as when groups are equally-weighted (e.g., $p_j = 1/m$). In the
latter case, only a weak decomposition of the aggregate RI and SRI
holds. The technical appendix also contains the derivation of the
designed-based standard errors for the total or aggregate RI and SRI
and their within-group components; see \citet{talih2012suppTA}.

\section{An entropy-based reference-invariant health disparity index}
\label{G}

Consider two nonnegative (yet, not necessarily probability) mass
functions $p$ and $q$. Suppose they are defined on a common set of
integers $\{1,2,\ldots,m\}$, which we take to be group membership
indicators for different socioeconomic and demographic groups in a
larger population. In analyses of health disparities, $p_j$ typically
denotes the relative population share of group $j$, whereas $q_j$
denotes its relative disease burden (or, inversely, the relative health
advantage). However, as discussed in Section \ref{intro}, other choices
for the quantities $p_j$ and $q_j$ may be desired. From the
mathematical point of view, investigating health disparities within the
population amounts to ascertaining the discrepancy between the two
distributions $p$ and $q.$

\begin{definition*} Based on a divergence proposed by \citet
{fujisawaeguchi2008} for robust parameter estimation in the presence
of outliers, and for $r_j = q_j/p_j$ and a scalar $\alpha\neq0, 1$,
\citet{cichockiamari2010} define the generalized R\'{e}nyi (or
alpha--gamma) divergence as
%
\begin{equation}
\label{G-ratio} R_{\alpha}(p\parallel  q) = \frac{1}{\alpha(1-\alpha)} \ln \biggl\{
\frac{
(\sum_{j=1}^m p_j )^\alpha (\sum_{j=1}^m p_jr_j )^{1-\alpha
}}{\sum_{j=1}^m p_j r_j^{1-\alpha}} \biggr\}.
\end{equation}
\end{definition*}

\begin{remark*}
Our approach, reflected in (\ref{G-ratio}) and
throughout the paper, differs from the standard information theoretic
approach in that we introduce dependence between the distributions $p$
and $q.$ The former is a weighting distribution---typically, the $p_j$
are the relative sizes of groups in the population. The latter is
constructed from $q_j = p_jr_j.$ Each $r_j$ specifies the disparity for
group $j$ relative to a common reference point, as explained in Section~\ref{common-practice}.
\end{remark*}
\subsection{Scale invariance and relation to R\'{e}nyi divergence}
\label{G1}

Due to the form of the argument of the logarithm in (\ref{G-ratio}),
the generalized R\'{e}nyi divergence $R_{\alpha}(p\parallel  q)$ is invariant to
rescaling of either the $p$ or the $q$ distributions. Indeed, for any
positive scalars $c_1$ and $c_2$,
\[
R_{\alpha}(c_1p\parallel  c_2q) = R_{\alpha}(p\parallel  q).
\]
In particular, for $\bar{p}_j = p_j/\sum p$, $\bar{q}_j = q_j/\sum q$,
and $\bar{r}_j = \bar{q}_j/\bar{p}_j$ we have
%
\begin{equation}
\label{renyi} R_{\alpha}(p\parallel  q) = R_{\alpha}(\bar{p}\parallel  \bar{q}) = -
\frac{1}{\alpha
(1-\alpha)} \ln \Biggl\{ \sum_{j=1}^m
\bar{p}_j \bar{r}_j^{1-\alpha} \Biggr\}.
\end{equation}
When $\alpha> 0$, the divergence $\alpha R_{\alpha}(\bar{p}\parallel  \bar{q})$
is the R\'{e}nyi divergence between two probability mass
functions---here, $\bar{p}$ and $\bar{q}$---introduced by \citet{renyi1960}.

\textit{Nonnegativity}. When $\alpha>0$ and $\alpha\neq1$,
Jensen's inequality ensures that $R_{\alpha}(\bar{p}\parallel  \bar{q}) \geq0$,
with equality if and only if $p= cq$ for some positive scalar $c;$ see,
for example, \citeauthor{vanErven2010} [(\citeyear{vanErven2010});
Chapter 6]. By skew-symmetry [see (\ref{skew-sym}) below], it follows
that $R_{\alpha}(p\parallel  q)\geq0$ for all $\alpha\neq0, 1.$

\textit{Monotonicity}. Jensen's inequality also ensures that the R\'
{e}nyi divergence $\alpha R_{\alpha}(p\parallel  q)$ is nondecreasing when $\alpha
>0$ increases; see \citeauthor{vanErven2010} (Chapter~6). Thus, when
$\alpha> 0$, $\alpha$ can be looked upon as an inequality (or
divergence) aversion parameter for the R\'{e}nyi divergence.

\textit{Practical relevance of scale invariance}. Henceforth, we
refer to the HDI that is derived from (\ref{renyi}) as the R\'{e}nyi
index ($\operatorname{RI}_{\alpha}$, or RI, for short). Scale invariance is
appropriate when it is believed that uniform proportional changes
across the population should leave the HDI unchanged; see \citet
{levyetal2006}. Scale invariance is especially desirable when seeking
HDIs that are invariant to the choice of the reference for evaluating
disparities, because, as seen in the \textit{Healthy People 2010 Final
Review}, identification of a reference group can be affected by
statistical reliability. In this respect, the RI remains the same,
whether we use the population average as the denominator for the
relative disparities $r_j$, as in (\ref{intro-rj}), the group with the
least adverse health outcome, as in (\ref{intro-rj2}), or take any
pre-set (positive) target, for example, a HP2010 or HP2020 target.

\subsection{Limiting cases} \label{G0}

The generalized R\'{e}nyi divergence is extended by continuity to the
limiting cases $\alpha\rightarrow1$ and $\alpha\rightarrow0$ (l'H\^
opital's rule):
%
\begin{equation}
\label{MLD-TI} R_1(p\parallel  q):= -\sum_{j=1}^m
\bar{p}_j\ln\bar{r}_j \quad\mbox{and}\quad R_0(p\parallel  q):= \sum_{j=1}^m \bar{p}_j\bar{r}_j\ln\bar{r}_j.
\end{equation}
When $p_j = n_j/n$, $q_j=p_jr_j$, and the $r_j$ are as in (\ref
{intro-rj}), these special limiting cases of the RI with $\alpha
\rightarrow1$ and $\alpha\rightarrow0$ are the MLD and the TI,
respectively; see \citet{borrelltalih2011}.

\textit{Interpretation of the MLD and the TI}. The MLD and the TI
were originally proposed as measures of income inequality by \citet
{theil1967}. Both the MLD and the TI are well-established measures of
relative entropy between two probability distributions, due to \citet
{kullbackleibler1951}. The general form of the Kullback--Leibler (K--L)
divergences is
%
\begin{eqnarray}
\label{KL-general} \operatorname{KL}(p\parallel  q) &=& \sum_{j=1}^m
p_j(r_j-1-\ln r_j) \quad\mbox{and}
\nonumber
\\[-8pt]
\\[-8pt]
\nonumber
\operatorname{KL}(q\parallel  p) &= &\sum_{j=1}^m
p_j(1-r_j+r_j\ln r_j).
\end{eqnarray}
When $p_j = n_j/n$, $q_j=p_jr_j$, and the $r_j$ are as in (\ref
{intro-rj}), $\operatorname{MLD} = \operatorname{KL}(p\parallel  q)$, whereas $\operatorname{TI}=\operatorname
{KL}(q\parallel  p).$ Thus, the MLD and the TI summarize the
disproportionalities between the relative sizes of groups in the
population and those groups' shares of an adverse health outcome. In
this regard, from (\ref{MLD-TI}), the MLD is seen as a log-likelihood
ratio test statistic for the null hypothesis that group shares of the
adverse health outcome have been ``allocated'' according to the relative
sizes of the groups in the population. Similarly, the TI tests the null
hypothesis that group shares of the total population have been
``allocated'' according to the groups' shares of the adverse health
outcome. This interpretation of the MLD and TI as log-likelihood ratio
tests will be revisited in the case study of Section \ref{case-studies}
to assess the statistical significance of the symmetrized R\'{e}nyi index.

\subsection{Symmetrized R\'{e}nyi index} \label{G3}

Generalized R\'{e}nyi divergence in (\ref{renyi}) is asymmetric in the
two distributions whose generalized relative entropy is being
evaluated: $R_{\alpha}(p\parallel  q)$ will be mostly influenced by groups with
large values of $p_j$, whereas $R_{\alpha}(q\parallel  p)$ will be mostly
influenced by groups with large values of $q_j.$ \citet
{borrelltalih2011} discuss this issue of lack of symmetry in the
context of the special cases $R_1(\bar{p}\parallel  \bar{q}) = \operatorname{KL}(\bar
{p}\parallel  \bar{q})$ (MLD) and $R_0(\bar{p}\parallel  \bar{q}) = \operatorname{KL}(\bar
{q}\parallel  \bar{p})$ (TI), with $p_j = n_j/n$, $q_j=p_jr_j$, and the $r_j$ as
in (\ref{intro-rj}). Yet,
%
\begin{equation}
\label{skew-sym} R_{\alpha}(q\parallel  p) = R_{1-\alpha}(p\parallel  q).
\end{equation}
Thus, a symmetrized generalized R\'{e}nyi divergence, ${SR}_{\alpha}(p,q)$,
is obtained from $ [R_{\alpha}(p\parallel  q) + R_{1-\alpha}(p\parallel  q) ]
/2.$ For $\bar{p}_j = p_j/\sum p$, $\bar{q}_j = q_j/\sum q$, and $\bar
{r}_j = \bar{q}_j/\bar{p}_j$, ${SR}_{\alpha}(p,q)$ is given by
%
\begin{equation}
\label{SRI} {SR}_{\alpha}(p, q) = {SR}_{\alpha}(\bar{p}, \bar{q}) = -
\frac{1}{2\alpha
(1-\alpha)} \ln \Biggl\{ \Biggl( \sum_{j=1}^m
\bar{p}_j \bar {r}_j^{1-\alpha} \Biggr) \Biggl(\sum
_{j=1}^m \bar{p}_j \bar
{r}_j^{\alpha} \Biggr) \Biggr\}.\hspace*{-35pt}
\end{equation}
We refer to the HDI that is derived from (\ref{SRI}) as the symmetrized
R\'{e}nyi index (SRI).

\textit{Limiting case}. As in Section \ref{G0}, ${SR}_{\alpha}(p, q)$ is
extended by continuity to the cases $\alpha\rightarrow1, 0$:
%
\begin{equation}
\label{STI} {SR}_1(p, q):= \frac{1}{2} \sum _{j=1}^m \bar{p}_j(
\bar{r}_j - 1)\ln\bar{r}_j =: {SR}_0(p, q).
\end{equation}
The divergence in (\ref{STI}) is a symmetrized Kulback--Leibler
divergence, also known as half the Jeffrey's divergence; see, for
example, \citet{pollard2002}. When $p_j=n_j/n$ is the population share
for group $j$, the $r_j$ are as in~(\ref{intro-rj}), and $q_j=p_jr_j$
is the disease share $y_{\cdot j}/y_{\cdot\cdot}.$ \citet
{borrelltalih2011} coin the symmetrized divergence in~(\ref{STI}) the
symmetrized Theil index (STI).

\subsection{Standardization and relation to the Atkinson index} \label{G2}

For $\alpha> 0$, a~standardized generalized R\'{e}nyi divergence, with
values between 0 and 1, and which we denote by $A_{\alpha}(p\parallel  q)$, can be
defined for any nonnegative (not necessarily probability)
distributions $p$ and $q$:
%
\begin{equation}
\label{exp-transform} A_{\alpha}(p\parallel  q) = 1-e^{-\alpha R_{\alpha}(p\parallel  q)}.
\end{equation}
Thus, when $\alpha> 0$, we have
\[
A_{\alpha}(p\parallel  q) = 1 - \Biggl( \sum_{j=1}^m
\bar{p}_j \bar{r}_j^{1-\alpha
} \Biggr)^{{1}/{(1-\alpha)}}.
\]
For $p_j = n_j/n$, $q_j=p_jr_j$, and the $r_j$ as in (\ref{intro-rj}),
this is the (between-group) Atkinson index, introduced by \citet
{atkinson1970} for measuring income inequalities, with parameter
$\alpha> 0$ quantifying society's aversion to inequality.\vadjust{\goodbreak}

\textit{Standardized SRI}. Applying a standardizing exponential
transformation similar to the one in (\ref{exp-transform}), we
construct a standardized SRI, with values between 0 and 1, as follows:
%
\begin{equation}
\label{standard-SRI} SA_{\alpha}(p, q) = %
\cases{
1-e^{-\alpha {SR}_{\alpha}(p, q)},&\quad $\mbox{when } \alpha\geq1/2,$
\vspace*{2pt}\cr
1-e^{-(1-\alpha) {SR}_{\alpha}(p, q)},&\quad $\mbox{when } \alpha<
1/2.$}
\end{equation}
This construction preserves symmetry of the SRI around the parameter
value $\alpha= 1/2.$ Since $\alpha {SR}_{\alpha}(p,q)$ is nondecreasing
for $\alpha\geq1/2$, $\alpha$ is a disparity aversion parameter for
the standardized SRI. By symmetry, $1-\alpha$ is a disparity aversion
parameter when $\alpha< 1/2.$ The value $\alpha= 1/2$ can be
interpreted as the most conservative choice for disparity aversion in
the standardized SRI, in that it gives a lower bound for the index.

\subsection{The RI and SRI as between-group HDIs}\label{SRIb}

By construction, we have $q_j = p_j r_j$ and, from (\ref{intro-rj}) or
(\ref{intro-rj2}), $r_j \propto\bar{y}_{\cdot j}.$
From (\ref{G-ratio}) and (\ref{SRI}), we have expressions for the
between-group RI and SRI in terms of the group sizes $n_j$
and means $\bar{y}_{\cdot j}$, which we list next for $\alpha\neq0,1.$
Henceforth, to distinguish the between-group RI
(resp., SRI) from the within-group RI (resp., SRI) and
the aggregate or total RI (resp., SRI)
that are discussed in the technical appendix [\citeauthor{talih2012suppTA} (\citeyear{talih2012suppTA})], we
use the notation $[\operatorname{RI}]_{\mathrm{B}}$ (resp., $[\operatorname
{SRI}]_{\mathrm{B}}$):

\begin{itemize}
\item Population-weighted group contributions $p_j = n_j/n$,
%
\begin{eqnarray}
{[\operatorname{RI}_{\alpha}]}_{\mathrm{B}} &=& \frac{1}{\alpha(1-\alpha)} \ln \biggl\{
\frac{n^\alpha (\sum_{j=1}^m n_j\bar{y}_{\cdot j}
)^{1-\alpha}}{\sum_{j=1}^m n_j \bar{y}_{\cdot j}^{1-\alpha}} \biggr\}, \label{PWRI}
\\
{[\operatorname{SRI}_{\alpha}]}_{\mathrm{B}} &=& \frac{1}{2\alpha(1-\alpha)} \ln \biggl\{
\frac{n \sum_{j=1}^m n_j\bar{y}_{\cdot j}}{ (\sum_{j=1}^m n_j \bar{y}_{\cdot j}^{1-\alpha} )  (\sum_{j=1}^m n_j
\bar{y}_{\cdot j}^{\alpha} )} \biggr\}. \label{PWSRI}
\end{eqnarray}
\item Equally-weighted group contributions $p_j = 1/m$,
%
\begin{eqnarray}
{\bigl[\operatorname{RI}'_{\alpha}\bigr]}_{\mathrm{B}} &=&
\frac{1}{\alpha(1-\alpha)} \ln \biggl\{ \frac{m^\alpha (\sum_{j=1}^m \bar{y}_{\cdot j}
)^{1-\alpha}}{\sum_{j=1}^m \bar{y}_{\cdot j}^{1-\alpha}} \biggr\}, \label {EWRI}
\\
{\bigl[\operatorname{SRI}'_{\alpha}\bigr]}_{\mathrm{B}} &=&
\frac{1}{2\alpha(1-\alpha)} \ln \biggl\{\frac{m \sum_{j=1}^m \bar{y}_{\cdot j}}{ (\sum_{j=1}^m
\bar{y}_{\cdot j}^{1-\alpha} )  (\sum_{j=1}^m \bar{y}_{\cdot
j}^{\alpha} )} \biggr\}. \label{EWSRI}
\end{eqnarray}

\end{itemize}

\textit{Limiting cases}. The expressions for the RI and SRI when
$\alpha\rightarrow1$ or $\alpha\rightarrow0$ are obtained by taking
limits in (\ref{PWRI})--(\ref{EWSRI}) above. We list them here for ease
of reference. Equation~(\ref{PWRI}) with $\alpha\rightarrow1$ yields the MLD,
\[
[\operatorname{RI}_1]_{\mathrm{B}}:= - \frac{1}{n}\sum_{j=1}^m n_j \ln\bar {y}_{\cdot j}+ \ln\bar{y}_{\cdot\cdot},
\]
whereas $\alpha\rightarrow0$ yields the TI,
\[
[\operatorname{RI}_0]_{\mathrm{B}}:= \frac{1}{n\bar{y}_{\cdot\cdot}}\sum
_{j=1}^m n_j \bar{y}_{\cdot j} \ln
\bar{y}_{\cdot j} - \ln\bar {y}_{\cdot\cdot}.
\]
As well, (\ref{PWSRI}) with either $\alpha\rightarrow1$ or $\alpha
\rightarrow0$ yields the STI,
\[
[\operatorname{SRI}_1]_{\mathrm{B}}:= \frac{1}{2n\bar{y}_{\cdot\cdot}}\sum_{j=1}^m n_j (\bar{y}_{\cdot j} -
\bar{y}_{\cdot\cdot}) \ln\bar {y}_{\cdot j} =: [\operatorname{SRI}_0]_{\mathrm{B}}.
\]
On the other hand, taking the limit when $\alpha\rightarrow1$ in (\ref
{EWRI}) results in
\[
\bigl[\operatorname{RI}'_1\bigr]_{\mathrm{B}}:= -
\frac{1}{m}\sum_{j=1}^m \ln\bar
{y}_{\cdot j} + \ln \Biggl[\frac{1}{m}\sum_{j=1}^m \bar{y}_{\cdot
j} \Biggr],
\]
while the limit when $\alpha\rightarrow0$ is
\[
\bigl[\operatorname{RI}'_0\bigr]_{\mathrm{B}}:=
\frac{1}{\sum_{k=1}^m \bar{y}_{\cdot
k}} \sum_{j=1}^m
\bar{y}_{\cdot j} \ln\bar{y}_{\cdot j} - \ln \Biggl[\frac{1}{m}
\sum_{j=1}^m \bar{y}_{\cdot j}
\Biggr].
\]
Thus, the limit in (\ref{EWSRI}) when $\alpha\rightarrow1$ or $0$ is
\[
\bigl[\operatorname{SRI}'_1\bigr]_{\mathrm{B}}:=
\frac{1}{2\sum_{k=1}^m \bar{y}_{\cdot
k}} \sum_{j=1}^m \Biggl[
\bar{y}_{\cdot j} - \frac{1}{m} \sum_{k=1}^m
\bar{y}_{\cdot k} \Biggr]\ln\bar{y}_{\cdot j} =: \bigl[\operatorname
{SRI}'_0\bigr]_{\mathrm{B}}.
\]

\section{Alternatives to the R\'{e}nyi index}\label{alternatives}

\subsection{Generalized entropy class} \label{ALPHA}

A class of measures that originate in the measurement of income
inequalities is the generalized entropy (GE) class, which specifies
$p_j = n_j/n$ and the ratios $r_j$ as in (\ref{intro-rj}); see \citet
{biewenjenkins2006}, \citet{elbersetal2008}, and references therein.
The GE class is a special case of alpha divergence. The latter was
introduced by \citet{chernoff1952} to evaluate the asymptotic
efficiency of likelihood ratio tests. \citet{cressieread1984} also
discuss such measures for multinomial goodness-of-fit tests. Using the
parameterization in \citet{cichockiamari2010}, alpha divergence is
defined for any nonnegative mass functions $p$ and $q$ and any real
number $\alpha$, $\alpha\neq0,1$, as
%
\begin{equation}
\label{alpha-ratios} D_{\alpha}(p\parallel  q) = \frac{1}{\alpha(1-\alpha)} \sum
_{j=1}^m p_j \bigl[\alpha+ (1-\alpha)
r_j - r_j^{1-\alpha}\bigr],
\end{equation}
where, as before, the $r_j$ are the ratios $r_j = q_j / p_j.$ Just like
the generalized R\'{e}nyi divergence, $D_{\alpha}(p\parallel  q)$ can be extended
by continuity to the limiting cases $\alpha\rightarrow1$ and $\alpha
\rightarrow0$, yielding the K--L divergences in (\ref{KL-general}). It
is well known that alpha divergence\vadjust{\goodbreak} $D_{\alpha}(p\parallel  q)$ remains
nonnegative, $D_{\alpha}(p\parallel  q) \geq0$, with equality if and only if
$p_j=q_j$ for each $j$ in $1,2,\ldots, m.$ When $p$ and $q$ are
probability mass functions, that is, $\sum_{j=1}^m p_j = 1$ and $\sum_{j=1}^m q_j =1$, alpha divergence is a Csisz\'{a}r $f$-divergence; see
\citet{alisilvey1966}.

We refer to the index that is derived from (\ref{alpha-ratios}) as the
GE index. Just like with the R\'{e}nyi index, a symmetrized GE index is
obtained simply by taking the arithmetic average of $D_{\alpha}(p\parallel  q)$
and $D_{1-\alpha}(p\parallel  q)$:
%
\begin{equation}
\label{sym-alpha-ratios} SD_{\alpha}(p, q) = \frac{1}{2\alpha(1-\alpha)} \sum
_{j=1}^m p_j \bigl(1 +
r_j - r_j^{1-\alpha} - r_j^\alpha
\bigr).
\end{equation}
In addition, the symmetrized GE index in (\ref{sym-alpha-ratios}) can
be standardized to take values between 0 and 1 using the exponential
transformation in (\ref{standard-SRI}). Further, whereas alpha
divergence is not scale-invariant---it only holds that, for a positive
scalar~$c$, $D_{\alpha}(cp\parallel  cq) = c D_{\alpha}(p\parallel  q)$---a
reference-invariant GE index $D_{\alpha}(\bar{p}\parallel  \bar{q})$ can be
constructed easily using the normalized distributions $\bar{p}$ and
$\bar{q}$, because $\overline{c_1p} = \bar{p}$ and $\overline{c_2q} =
\bar{q}$ for any positive scalars $c_1$ and $c_2.$ (As before, $\bar
{p}_j = p_j/\sum p$, $\bar{q}_j = q_j/\sum q$, and $\bar{r}_j = \bar
{q}_j/\bar{p}_j.$) Thus,
%
\begin{equation}
\label{ref-inv-GE} SD_{\alpha}(\bar{p}, \bar{q}) = \frac{1}{\alpha(1-\alpha)}
\Biggl[ 1- \frac{1}{2} \sum_{j=1}^m
\bar{p}_j \bigl(\bar{r}_j^{1-\alpha} + \bar
{r}_j^\alpha\bigr) \Biggr].
\end{equation}
We refer to this HDI as the symmetrized reference-invariant GE index.

\begin{proposition*} For nonnegative mass functions $p$ and $q$ on
$\{1, 2, \ldots, m\}$, let $\bar{p}_j = p_j/\sum p$, $\bar{q}_j =
q_j/\sum q$, and $\bar{r}_j = \bar{q}_j/\bar{p}_j.$ For ${SR}_{\alpha}(\bar
{p}, \bar{q})$ in (\ref{SRI}) and $SD_{\alpha}(\bar{p}, \bar{q})$ in~(\ref
{ref-inv-GE}):
%
\begin{eqnarray}
\alpha {SR}_{\alpha}(\bar{p}, \bar{q}) &\leq&\alpha SD_{\alpha}(\bar{p},
\bar {q})\qquad \mbox{when } \alpha> 1; \label{SRIvsSGE}
\\
\alpha {SR}_{\alpha}(\bar{p}, \bar{q}) &\geq&\alpha SD_{\alpha}(\bar{p},
\bar {q})\qquad \mbox{when } \alpha< 1 \mbox{ and } \alpha\neq0;
\end{eqnarray}
with equality when $\alpha\rightarrow1$ or $\alpha\rightarrow0.$
\end{proposition*}
\begin{pf}
Without loss of generality, let $\alpha> 1.$ The
proof follows from the application of the arithmetic-geometric mean
inequality and the fact that $x-1 \geq\ln x$ for all $x >0.$
\end{pf}

In Section \ref{sim-studies} we show not only that the SRI is more
conservative than the symmetrized reference-invariant GE index for
$\alpha> 1$, as implied by (\ref{SRIvsSGE}), but also that the SRI is
more robust to small changes in the disease distribution $q$, which
renders it a more desirable HDI.

The GE class is well-studied in the economics literature. The GE class
is consistent with a certain set of axiomatic properties that are
relevant for income distributions; see, for example, \citet
{cowelletal2011}, \citet{cowellkuga1980}, and \citet{shorrocks1980}.
Even though such axioms are\vadjust{\goodbreak} not sufficient for health benefits
analyses, the GE class remains a widely used class for constructing
HDIs; see \citet{levyetal2006}. In addition to the K--L divergences
($\alpha\rightarrow1$ or $0$), special cases of alpha divergence in
(\ref{alpha-ratios}) are the Pearson ($\alpha=-1$) and Neyman ($\alpha
=2$) chi-squared statistics and the squared Hellinger distance ($\alpha
=0.5$). 
\subsection{Bregman class} \label{BETA}

Bregman divergences are generated from any twice differentiable and
strictly convex function $\Phi$ as follows:
\[
\label{bregman} B_\Phi(p\parallel  q) = \sum_{j=1}^m
\bigl[\Phi(q_j) - \Phi(p_j) -(q_j-p_j)
\Phi'(p_j)\bigr].
\]
A common choice for the generating function $\Phi$, for $\beta\neq
0,1$, is
\[
\label{Phi} \Phi(u) = \frac{\beta+ (1-\beta) u - u^{1-\beta}}{\beta(1-\beta)},
\]
which yields the beta divergence, defined for $\beta\neq0, 1$ and
$r_j=q_j/p_j$,
\begin{equation}
\label{beta-ratios} B_\beta(p\parallel  q) = \frac{1}{\beta(1-\beta)} \sum
_{j=1}^m p_j^{1-\beta} \bigl[
\beta+ (1-\beta) r_j - r_j^{1-\beta}\bigr],
\end{equation}
and appropriate extensions by continuity when $\beta\rightarrow0$ or
$1$; see \citet{cichockiamari2010}. As before, the limiting case $\beta
\rightarrow0$ reduces to the Kulback--Leibler divergence $\operatorname
{KL}(q\parallel  p)$ in (\ref{KL-general}). However, the case $\beta\rightarrow
1$ is no longer $\operatorname{KL}(p\parallel  q)$, but, instead, the so-called
Itakura--Saito (IS) divergence, given by\looseness=-1
\[
\operatorname{IS}(p\parallel  q) = \sum_{j=1}^m
(r_j - 1 - \ln r_j).
\]\looseness=0

Beta divergence in (\ref{beta-ratios}) provides a class of HDIs that
are worth investigating in future work. For instance, a symmetrized
reference-invariant beta divergence is obtained from $[B_\beta(p\parallel  q) +
B_\beta(q\parallel  p)]/2$, resulting in
\[
SB_\beta(\bar{p}, \bar{q}) = -\frac{1}{2\beta}\sum
_{j=1}^m \bar {p}_j^{1-\beta} (1-
\bar{r}_j) \bigl(1-\bar{r}_j^{-\beta}\bigr).
\]
However, as explained in Section \ref{intro}, the $p_j$ are weights
that are assigned by the analyst to each population group $j$, commonly
using either equal weights (e.g., $1/m$) or size-based weights (e.g.,
$n_j/n$). Therefore, in the context of this paper, the analyst would
need to provide additional justification for the logarithmic rescaling
of the $p_j$ in (\ref{beta-ratios}) by the factor $1-\beta$, which is
not the case for alpha divergence~(\ref{alpha-ratios}) or generalized
R\'{e}nyi divergence (\ref{renyi}).

\citet{magdalounock2011} derive the Bregman class as the unique class
of measures that are consistent with certain inequality measurement
principles, including the transfer principle (albeit modified) and
decomposability. By the authors' own assessment, the key to their
derivation is a new principle of ``judgment separability'' that they
introduce for the analysis of income inequalities. In this paper, we
restrict attention to reference-invariant HDIs (i.e., strong
scale-invariant measures), whereby judgment separability is not
necessary, because it reduces to the weaker principle of
``indiscernability of identicals.'' The latter postulates simply that an
inequality measure $D$ satisfies $D(p\parallel  p) = 0$ for any distribution $p$.

For those reasons, we do not discuss beta divergence in Section \ref
{sim-studies}; we compare the SRI only with the symmetrized
reference-invariant GE index.

\subsection{SRI and changes therein}\label{sim-studies}

To illustrate the robustness of the SRI (\ref{SRI}) to small changes in
the distribution $q=pr$, for a fixed $p$, and in comparison with the
symmetrized reference-invariant GE index in (\ref{ref-inv-GE}), we
examine the SRI under simple scenarios borrowed from \citet{harperetal2010}.
%

\begin{table}
\caption{Baseline and three hypothetical
scenarios to examine changes in the SRI and its GE-based counterpart}\label{case-studies-tab}
\begin{tabular*}{\textwidth}{@{\extracolsep{\fill}}lcccc@{}}
\hline
\multicolumn{1}{@{}l}{\textbf{Group} $\bolds{(j)}$} &\textbf{A} &\textbf{B} &\textbf{C} &\textbf{D}\\
\hline
Relative size $n_j/n$ &25\% &25\% &25\% &25\%\\
\textit{Baseline} & & & &\\
\quad Group rate $\bar{y}_{\cdot j}$ &50\% &40\% &30\% &10\%\\
\textit{Scenario} 1 & & & &\\
\quad Group rate $\bar{y}_{\cdot j}$ &50\% &\textbf{30\%} &30\% &10\%\\
\textit{Scenario} 2 & & & &\\
\quad Group rate $\bar{y}_{\cdot j}$ &\textbf{40\%} &40\% &30\% &10\%\\
\textit{Scenario} 3 & & & &\\
\quad Group rate $\bar{y}_{\cdot j}$ &50\% &40\% &\textbf{40\%} &10\%\\
\hline
\end{tabular*}
\end{table}

\begin{figure}

\includegraphics{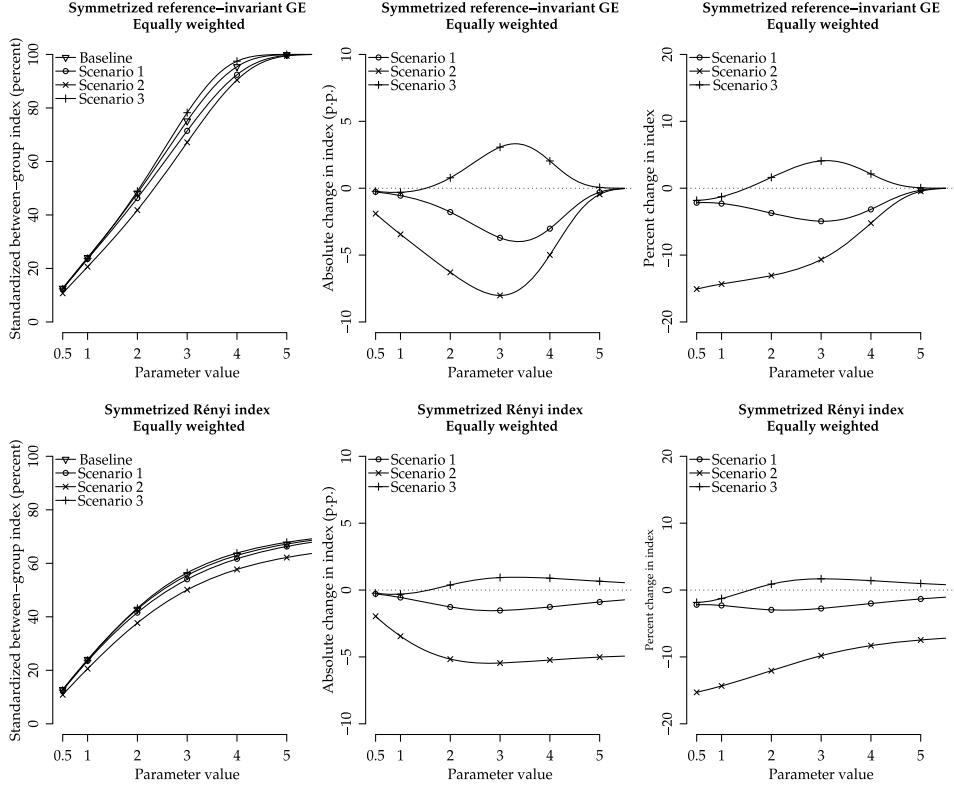}

\caption{Comparison of the symmetrized R\'{e}nyi index (SRI) with the
symmetrized reference-invariant GE index. The top- and bottom-left
panels show the symmetrized reference-invariant GE and the SRI,
respectively, under the scenarios described in Table \protect\ref
{case-studies-tab} for different values of the disparity aversion
parameter $\alpha.$ Only values of $\alpha\geq0.5$ are shown due to
symmetry. As confirmed in (\protect\ref{SRIvsSGE}), the standardized SRI is
more conservative for parameter values $\alpha> 1.$ In addition, the
SRI is seen to better discriminate between the different scenarios in
Table \protect\ref{case-studies-tab} for large values of $\alpha.$ Furthermore,
for $\alpha\leq3$ (approximately), the change in the SRI is
considerably smaller than the change in the symmetrized
reference-invariant GE, both in the absolute as well as the relative
scales, which illustrates the robustness of the SRI to small changes in
the distribution of the adverse health outcome.} \label{case-studies-ew}
\end{figure}

In Table \ref{case-studies-tab} we consider a population that is
divided into four groups of equal size, so that the population-weighted
distribution $p_j=n_j/n$ is the same as the equally-weighted
distribution $p_j=1/4.$ At baseline, group D has the least adverse
health outcome, with a rate of 10\%, whereas group A has the most
adverse outcome, with a rate of 50\%. Groups B and C have rates of 40\%
and 30\%, respectively. In scenarios 1 and 3, the groups with the least
and most adverse outcomes remain the same, but in scenario 1 the rate
for group B decreases 10 percentage points from baseline, whereas in
scenario 3 the rate for group C increases 10 percentage points from
baseline, in both scenarios achieving equal rates for groups~B and C.
In scenario 2, the rate for group A decreases 10 percentage points
while the other group rates remain unchanged. Because in scenario 3
group D (the ``best-off'' group) is further separated from the other
groups, with a 30 percentage points difference from the next best
(group C), compared to a 20 percentage points difference at baseline,\vadjust{\goodbreak}
we expect that disparities will increase overall. In scenario 2, the
gap between the best-off group (group~D) and the worst-off group (group~A)
has decreased, therefore, we expect an overall decrease in
disparities. In scenario~1, we similarly expect a decrease in
disparities because the rate for group B has moved closer to the best rate.

The top- and bottom-left panels in Figure \ref{case-studies-ew} compare
the symmetrized reference-invariant GE and the symmetrized R\'{e}nyi
indices under the above scenarios for different values of the disparity
aversion parameter $\alpha.$ Only values of $\alpha\geq0.5$ are shown
due to symmetry. As confirmed in (\ref{SRIvsSGE}), the standardized SRI
is seen to be more conservative for values of $\alpha> 1$; moreover,
the SRI is seen to better discriminate between the different scenarios
in Table \ref{case-studies-tab} for large values of $\alpha.$ Indeed,
observe how the symmetrized reference-invariant GE can no longer
distinguish between the various scenarios for large values of $\alpha$,
whereas the SRI still can. This is observed both in the absolute scale,
in the top- and bottom-center panels, as well as the relative scale, in
the top- and bottom-right panels. Furthermore, for $\alpha\leq3$
(approximately), the change in the SRI is considerably smaller than the
change in the symmetrized reference-invariant GE, both in the absolute
as well as in the relative scales, which illustrates the robustness of
the SRI to small changes in the distribution of the adverse health outcome.

\section{Design-based standard errors} \label{stderrors}

Mart{\'{\i}}nez-Camblor (\citeyear{martinez-camblor2007}) establishes a central limit theorem for
the total TI under simple random sampling. \citet{cowelletal2011} use
similar empirical processes techniques to analyze the asymptotic
distribution of goodness-of-fit statistics that are derived from the GE
class. Using Taylor series linearization, \citet{biewenjenkins2006}
derive the sampling variances for both the total TI and MLD---as well
as the GE class of total HDIs---for complex survey data. \citet
{borrelltalih2011} extend the Taylor series linearization method to
the case of grouped complex survey data to obtain the sampling variance
of the total STI and its between-group and within-group components.
\citet{borrelltalih2011} validate the sampling variances obtained via
linearization by comparing them to the ones obtained via balanced
repeated replication and rescaled bootstrap, which are developed in
\citet{mccarthy1969}, \citet{fay1989}, \citet{judkins1990}, \citet
{raowu1988}, \citet{raoetal1992}, and discussed in the context of
health inequality measures in \citet{harperetal2008} and \citet
{chengetal2008}. In this paper, we adopt a strategy similar to the
one in \citet{borrelltalih2011}, using Taylor series linearization,
balanced repeated replication, and the rescaled bootstrap to evaluate
and validate the design-based standard errors for the RI (and, by
extension, the SRI) and its between- and within- group components.
Below, we only show the calculations for the between-group component
$[\operatorname{RI}_{\alpha}]_{\mathrm{B}}.$ The calculations for the sampling
variance for the within-group component $[\operatorname{RI}_{\alpha}]_{\mathrm{W}}$
are shown in the technical appendix; see \citet{talih2012suppTA}. Also,
because $\operatorname{SRI}_{\alpha}= (\operatorname{RI}_{\alpha}+ \operatorname{RI}_{1-\alpha
})/2$, the sampling variance for the SRI and its between- and
within-group components easily follows. R code for computing the total
RI and SRI, together with their group-specific, between-, and
within-group components in grouped complex survey data, as well as
their design-based standard errors, is provided as a supplement; see
\citet{talih2012suppRR}.

Define, for any real number $a$,
%
\begin{eqnarray}
U_{a, j} &=& \sum_{s=1}^S \sum
_{c=1}^{C_s}\sum_{i=1}^{l_{cs}}
\delta _{icsj} w_{ics} y^a_{ics}, \label{Uj}
\\
U_{a,\cdot} &=& \sum_{j=1}^m
U_{a, j}.\label{Udot}
\end{eqnarray}
In the above, $S$ is the number of strata; $C_s$ is the number of PSU's
in stratum~$s$; $l_{cs}$ is the number of sample observations in the
PSU-stratum pair $(c,s);$ $w_{ics}$ is the sampling weight for sample
observation $i$ in the PSU-stratum pair $(c,s);$ $y_{ics}$ is the
severity of the adverse health outcome for sample observation $i$ in
the PSU-stratum pair $(c,s);$ $\delta_{icsj}=1$ when observation $i$
[in PSU-stratum pair $(c,s)$] belongs to group $j$ and $\delta_{icsj}
=0$ otherwise; and $j$ ranges from $1$ to $m$, where $m$ is the number
of groups in the population. With the notation introduced in (\ref{Uj})
and~(\ref{Udot}), we have $n = U_{0, \cdot}$, $n_j = U_{0, j}$,
$y_{\cdot\cdot} = U_{1, \cdot}$, $y_{\cdot j} = U_{1, j}$, $\bar
{y}_{\cdot\cdot} = U_{1, \cdot}/U_{0, \cdot}$, and $\bar{y}_{\cdot j}
= U_{1, j}/U_{0, j}.$

\subsection{Population-weighted groups: $p_j = n_j/n$}

From (\ref{PWRI}), we see that the between-group component $[\operatorname
{RI}_{\alpha}]_B$ can be written as a function solely of the sufficient
statistics $U_{a, k}$ in (\ref{Uj}). Thus, when $\alpha\neq0, 1$, the
partial derivatives with respect to $U_{0, k}$ and $U_{1, k}$ are
%
\begin{eqnarray}
\frac{\partial}{\partial U_{0, k}} [\operatorname{RI}_{\alpha}]_{\mathrm{B}} &=&
\frac{1}{1-\alpha} \biggl\{ \frac{1}{n} - \frac{\bar{y}_{\cdot
k}^{1-\alpha}}{\sum_{j=1}^m n_j \bar{y}_{\cdot j}^{1-\alpha}} \biggr\},
\label{G10k}
\\
\frac{\partial}{\partial U_{1, k}} [\operatorname{RI}_{\alpha}]_{\mathrm{B}} &=&
\frac{1}{\alpha} \biggl\{ \frac{1}{n\bar{y}_{\cdot\cdot}} - \frac{\bar{y}_{\cdot k}^{-\alpha}}{\sum_{j=1}^m n_j \bar{y}_{\cdot
j}^{1-\alpha}} \biggr\}.
\label{G11k}
\end{eqnarray}
The partial derivatives for the between-group component for the SRI,
which is given by $[\operatorname{SRI}_{\alpha}]_{\mathrm{B}} =  ([\operatorname
{RI}_{\alpha}]_{\mathrm{B}} + [\operatorname{RI}_{1-\alpha}]_{\mathrm{B}}
)/2$, easily follow.

\textit{Limiting cases}. When $p_j = n_j/n$, $r_j \propto\bar
{y}_{\cdot j}$, and $q_j = p_jr_j$, the distributions $\bar{p}=p/\sum p_j$ and $\bar{q} = q/\sum q_j$ are given by $\bar{p}_j = n_j/n$ and
$\bar{q}_j = \bar{p}_j\bar{r}_j$, respectively, with $\bar{r}_j = \bar
{y}_{\cdot j} /\bar{y}_{\cdot\cdot}$, as in (\ref{intro-rj}). Thus, the
limiting cases when $\alpha\rightarrow1$ or $0$ in~(\ref{G10k})--(\ref
{G11k}) reduce to the partial derivatives of the between-group MLD and
TI, respectively; see (\ref{MLD-TI}). These were computed in \citet
{borrelltalih2011}. We group them here for completeness:
\begin{eqnarray*}
\frac{\partial}{\partial U_{0, k}} [\operatorname{RI}_1]_{\mathrm{B}} &=&
\frac{1}{n^2}\sum_{j=1}^m
n_j \ln (\bar{y}_{\cdot j}/\bar {y}_{\cdot k} ),
\\
\frac{\partial}{\partial U_{1, k}} [\operatorname{RI}_1]_{\mathrm{B}} &=&
\frac{1}{n\bar{y}_{\cdot\cdot}} (1 - \bar{y}_{\cdot\cdot}/\bar {y}_{\cdot k}
),
\\
\frac{\partial}{\partial U_{0, k}} [\operatorname{RI}_0]_{\mathrm{B}} &=&
\frac{1}{n} (1- \bar{y}_{\cdot k}/\bar{y}_{\cdot\cdot} ),
\\
\frac{\partial}{\partial U_{1, k}} [\operatorname{RI}_0]_{\mathrm{B}} &=& -
\frac{1}{(n\bar{y}_{\cdot\cdot})^2}\sum_{j=1}^m
n_j\bar{y}_{\cdot j} \ln (\bar{y}_{\cdot j}/
\bar{y}_{\cdot k} ).
\end{eqnarray*}

Introduce an artificial variable $\sigma_{icsk}$ that represents the
variance contribution from each sample observation. The $\sigma_{icsk}$
are obtained by taking the dot product of the vector of partial
derivatives from (\ref{G10k})--(\ref{G11k}) with the vector of summands
in the sufficient statistics in (\ref{Uj}):
%
\begin{equation}
\label{sigmas} \sigma_{icsk} = \delta_{icsk} w_{ics}
\biggl\{\frac{\partial[\operatorname
{RI}_{\alpha}]_{\mathrm{B}}}{\partial U_{0, k}} + y_{ics} \frac{\partial
[\operatorname{RI}_{\alpha}]_{\mathrm{B}}}{\partial U_{1, k}} \biggr\}.
\end{equation}
Thus, an estimate of the sample variance of $[\operatorname{RI}_{\alpha}]_{\mathrm
{B}}$ is given by the sampling variance of the total statistic $\sum_{k=1}^m \sum_{i=1}^{l_{cs}} \sigma_{icsk}.$ The latter is readily
available, for example, using the command for survey estimation of
variances of totals (``svytotal'') in the R package ``survey''; see
\citeauthor{lumley2004} (\citeyear{lumley2004,Rsurvey2011}) and \citet{R2011}.

\subsection{Equally-weighted groups: $p_j = 1/m$}

From (\ref{EWRI}) and (\ref{Uj}), when $\alpha\neq0, 1$, the partial
derivatives with respect to $U_{0, k}$ and $U_{1, k}$ are given by
%
\begin{eqnarray}
\frac{\partial}{\partial U_{0, k}} \bigl[\operatorname{RI}'_{\alpha}
\bigr]_{\mathrm{B}} &=& -\frac{1}{\alpha  n_k} \biggl\{ \frac{\bar{y}_{\cdot k}}{\sum_{j=1}^m \bar{y}_{\cdot j}} -
\frac{\bar{y}_{\cdot k}^{1-\alpha}}{\sum_{j=1}^m \bar{y}_{\cdot j}^{1-\alpha}} \biggr\}, \label{G20k}
\\
\frac{\partial}{\partial U_{1, k}} \bigl[\operatorname{RI}'_{\alpha}
\bigr]_{\mathrm{B}} &=& \frac{1}{\alpha  n_k \bar{y}_{\cdot k}} \biggl\{ \frac{\bar
{y}_{\cdot k}}{\sum_{j=1}^m \bar{y}_{\cdot j}} -
\frac{\bar{y}_{\cdot
k}^{1-\alpha}}{\sum_{j=1}^m \bar{y}_{\cdot j}^{1-\alpha}} \biggr\}. \label{G21k}
\end{eqnarray}
The partial derivatives for the between-group component for the SRI,
which is given by $[\operatorname{SRI}'_{\alpha}]_{\mathrm{B}} =  ([\operatorname
{RI}'_{\alpha}]_{\mathrm{B}} + [\operatorname{RI}'_{1-\alpha}]_{\mathrm{B}}
)/2$, also follow.

\textit{Limiting cases}.
Limiting expressions for the partial derivatives in (\ref{G20k})--(\ref
{G21k}) of the between-group component $[\operatorname{RI}'_{\alpha}]_{\mathrm
{B}}$ are obtained as follows:
\begin{itemize}
\item When $\alpha\rightarrow1$,
\begin{eqnarray*}
\frac{\partial}{\partial U_{0, k}} \bigl[\operatorname{RI}'_1
\bigr]_{\mathrm{B}} &=& -\frac{1}{n_k} \biggl\{ \frac{\bar{y}_{\cdot k}}{\sum_{j=1}^m \bar
{y}_{\cdot j}} -
\frac{1}{m} \biggr\},
\\
\frac{\partial}{\partial U_{1, k}} \bigl[\operatorname{RI}'_1
\bigr]_{\mathrm{B}} &=& \frac{1}{n_k\bar{y}_{\cdot k}} \biggl\{ \frac{\bar{y}_{\cdot k}}{\sum_{j=1}^m \bar{y}_{\cdot j}} -
\frac{1}{m} \biggr\}.
\end{eqnarray*}

\item When $\alpha\rightarrow0$ (l'H\^{o}pital's rule),
\begin{eqnarray*}
\frac{\partial}{\partial U_{0, k}} \bigl[\operatorname{RI}'_0
\bigr]_{\mathrm{B}} &=& - \frac{\bar{y}_{\cdot k}}{n_k\sum_{j=1}^m \bar{y}_{\cdot j}} \biggl\{ \ln\bar{y}_{\cdot k}
- \frac{\sum_{j=1}^m \bar{y}_{\cdot j}\ln\bar
{y}_{\cdot j}}{\sum_{j=1}^m \bar{y}_{\cdot j}} \biggr\},
\\
\frac{\partial}{\partial U_{1, k}} \bigl[\operatorname{RI}'_0
\bigr]_{\mathrm{B}} &=& \frac{1}{n_k\sum_{j=1}^m \bar{y}_{\cdot j}} \biggl\{ \ln\bar {y}_{\cdot k} -
\frac{\sum_{j=1}^m \bar{y}_{\cdot j}\ln\bar{y}_{\cdot
j}}{\sum_{j=1}^m \bar{y}_{\cdot j}} \biggr\}.
\end{eqnarray*}

\end{itemize}

\section{Case study from NHANES}\label{case-studies}

HP2020 objective OH-5 in the Oral Health Topic Area aims to reduce the
proportion of U.S. adults aged 45--74 with moderate or severe
periodontitis. Table \ref{OHtab} presents estimated prevalence (and
standard errors) from NHANES 2001--04. The gradient associated with
socioeconomic status and the differences by sex and by race/ethnicity
are well documented; see, for example, \citet{borrelltalih2012}.

\begin{table}
\caption{Prevalence (in percent) of moderate or severe
periodontitis among U.S. adults aged 45--74, 2001--04\protect\tabnoteref{t1}}\label{OHtab}
\begin{tabular*}{\textwidth}{@{\extracolsep{\fill}}ld{2.1}d{1.3}d{2.1}d{2.1}@{}}
\hline
& \multicolumn{1}{c}{\textbf{Percent}} & \multicolumn{1}{c}{\textbf{SE}\tabnoteref{t2}} & \multicolumn{2}{c@{}} {\textbf{95\%
CI}\tabnoteref{t3}} \\
\hline
{Total} & 12.8 & 0.755 & 11.2 & 14.3 \\
{Sex} & & & & \\
\quad Male & 16.3 & 0.941 & 14.3 & 18.2 \\
\quad Female & 9.4 & 0.882 & 7.6 & 11.2 \\
{Race/Ethnicity} & & & & \\
\quad White only, non-Hispanic & 10.5 & 0.861 & 8.8 & 12.3 \\
\quad Black only, non-Hispanic & 22.1 & 1.863 & 18.3 & 25.9
\\
\quad Mexican-American & 18.1 & 2.829 & 12.3 & 23.9 \\
\quad Other\tabnoteref{t4} & 20.3 & 3.637 & 12.9 & 27.8\\
{Educational attainment} & & & & \\
\quad Less than high school & 26.8 & 1.974 & 22.8 & 30.9 \\
\quad High school graduate & 14.6 & 1.811 & 10.9 & 18.3 \\
\quad Some college or AA degree & 11.3 & 0.899 & 9.5 & 13.2\\
\quad College graduate or above & 6.5 & 1.191 & 4.0 & 8.9 \\
{Family income (percent FPL\tabnoteref{t5})} & & & & \\
\quad Less than 100 & 28.2 & 3.305 & 21.4 & 34.9 \\
\quad 100--199 & 24.1 & 2.134 & 19.7 & 28.4 \\
\quad 200--399 & 10.8 & 1.413 & 8.0 & 13.7 \\
\quad 400--499 & 8.4 & 1.504 & 5.3 & 11.4 \\
\quad 500 or above & 8.5 & 1.089 & 6.3 & 10.7 \\
\quad N/A\tabnoteref{t6} & 13.3 & 3.362 & 6.5 & 20.2 \\
{Country of birth} & & & & \\
\quad U.S. & 11.8 & 0.713 & 10.4 & 13.3 \\
\quad Outside U.S. & 19.2 & 2.622 & 13.9 & 24.6\\
\hline
\end{tabular*}
\tabnotetext[1]{t1}{Data are from the National Health and Nutrition Examination
Survey (NHANES) 2001--02 and 2003--04. The case definitions adopted by
the CDC working group for use in population-based surveillance of
periodontitis are as follows: for severe periodontitis, it is required
that two or more interproximal sites have clinical attachment loss
(CAL) $\geq$ 6~mm, not on the same tooth, and one or more interproximal
sites have pocket depth (PD) $\geq$ 5~mm; for moderate periodontitis, it
is required that either two or more interproximal sites have CAL $\geq$
4~mm, not on the same tooth, or two or more interproximal sites have PD
$\geq$ 5~mm, not on the same tooth. \citet{pageeke2007} explain the
rationale for those cutoff values.}
%
\tabnotetext[2]{t2}{Designed-based standard errors (SE) obtained via Taylor
linearization (e.g., SUDAAN or R ``survey'' package).}
\tabnotetext[3]{t3}{Lower and upper confidence limits, respectively, for a 95 percent
confidence interval (CI).}
\tabnotetext[4]{t4}{The category \textit{Other} consists of Hispanic or Latino other
than Mexican-American and non-Hispanic of races other than black and
white, including multiracial adults. The category \textit{Other} is listed
to provide a complete partition of the population into mutually
exclusive groups, but it is not part of the HP2020 population template
for objectives monitored using NHANES 1999 and later.}
\tabnotetext[5]{t5}{Family income as a percent of the federal poverty level (FPL),
also known as the poverty income ratio (PIR).}
\tabnotetext[6]{t6}{Adults whose family PIR is not available (N/A), listed to
maintain a complete partition of the population.}
\end{table}

Figure \ref{OHpanel} compares the standardized SRI for values of the
parameter $\alpha$ when groups are population-weighted and when groups
are equally-weighted. As seen in Section \ref{SRIb}, the
population-weighted SRI uses the estimated distributions (displayed in
the table within each figure panel) for the relative shares of
population ($p_j= n_j/n$) and of disease ($q_j = y_{\cdot j}/y_{\cdot
\cdot}$) in the symmetrized R\'{e}nyi divergence $S_{\alpha}(p, q)$,
whereas the equally-weighted SRI uses $p_j = 1/m.$ Due to symmetry of
the SRI around the parameter value $0.5$, only values of $\alpha\geq
0.5$ are shown. For values of $\alpha\geq0.5$, the parameter $\alpha$
is a disparity aversion parameter for the standardized SRI: the
standardized SRI is nondecreasing in $\alpha$ for $\alpha\geq0.5.$
The rescaled bootstrap method allows the design-based estimation of the
sampling distribution of the index. The box plots in Figure \ref
{OHpanel} represent the bootstrapped sampling distributions for the
different values of $\alpha$ and types of indices shown.

\begin{figure}
\centering
\begin{tabular}{@{}cc@{}}

\includegraphics{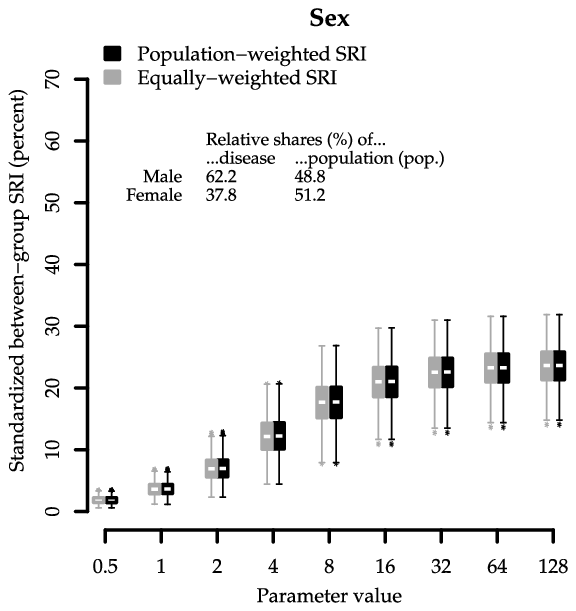}
 & \includegraphics{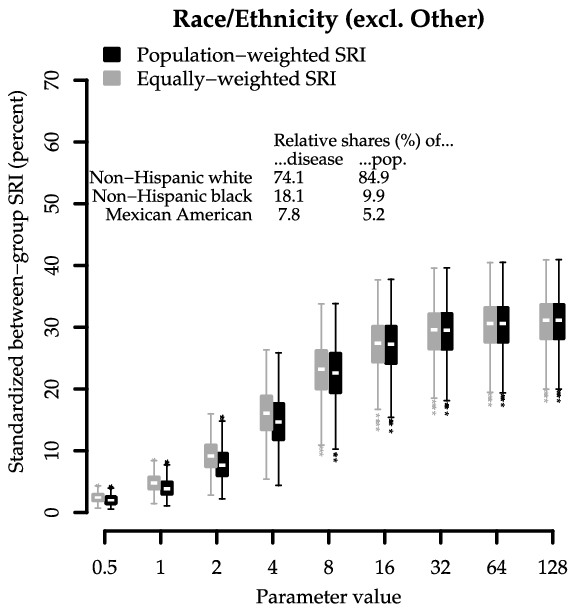}\\[3pt]

\includegraphics{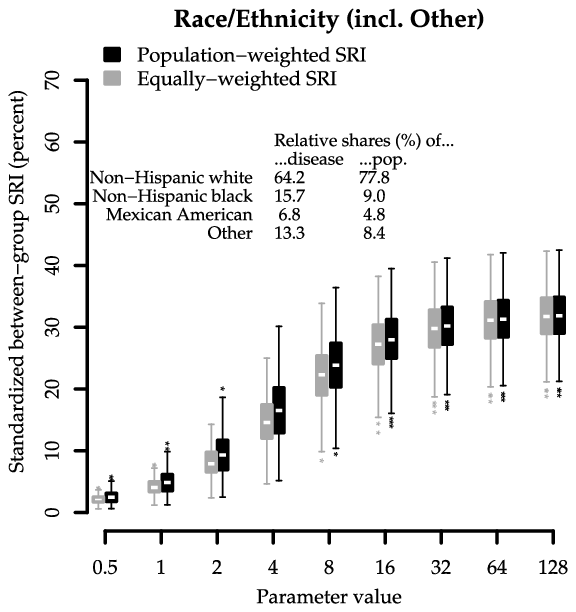}
 & \includegraphics{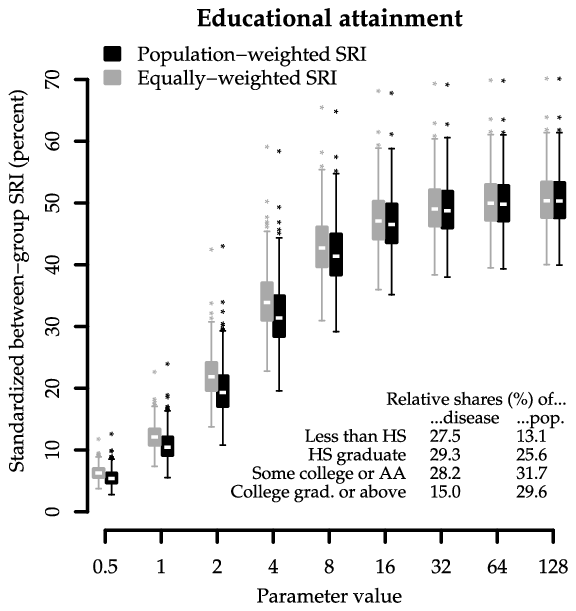}
\end{tabular}
\caption{Standardized between-group SRI by population characteristic
for the prevalence of moderate or severe periodontitis among U.S.
adults aged 45--74, 2001--04. Due to symmetry of the SRI around the
parameter value $0.5$, only values of $\alpha\geq0.5$ are shown. For
values of $\alpha\geq0.5$, the parameter $\alpha$ is a disparity
aversion parameter for the standardized SRI: the standardized SRI is
nondecreasing in $\alpha$ for $\alpha\geq0.5$; see (\protect\ref
{standard-SRI}). The population-weighted SRI uses the estimated
distributions (displayed in the table within each figure panel) for the
relative shares of population ($p_j= n_j/n$, restricting to individuals
with valid periodontal data) and of disease ($q_j = y_{\cdot j}/y_{\cdot
\cdot}$) in the symmetrized R\'{e}nyi divergence $S_{\alpha}(p, q)$,
whereas the equally-weighted SRI uses $p_j = 1/m.$ The box plots are
design-based, obtained via rescaled bootstrap with 500 replications.}
\label{OHpanel}
\end{figure}

\setcounter{figure}{1}
\begin{figure}
\centering
\begin{tabular}{@{}cc@{}}

\includegraphics{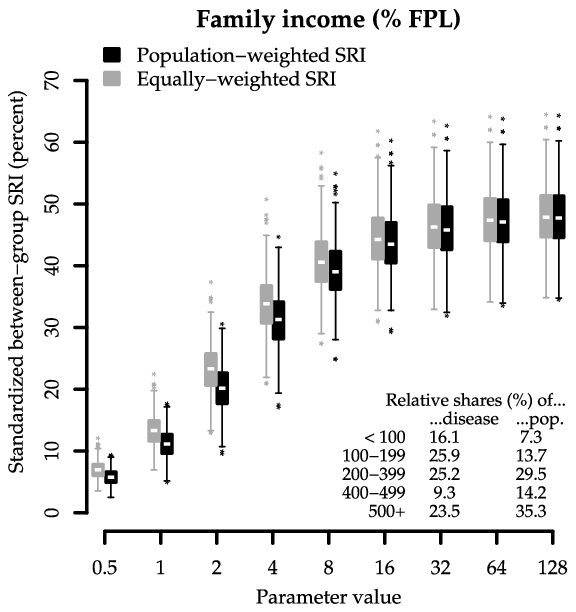}
 & \includegraphics{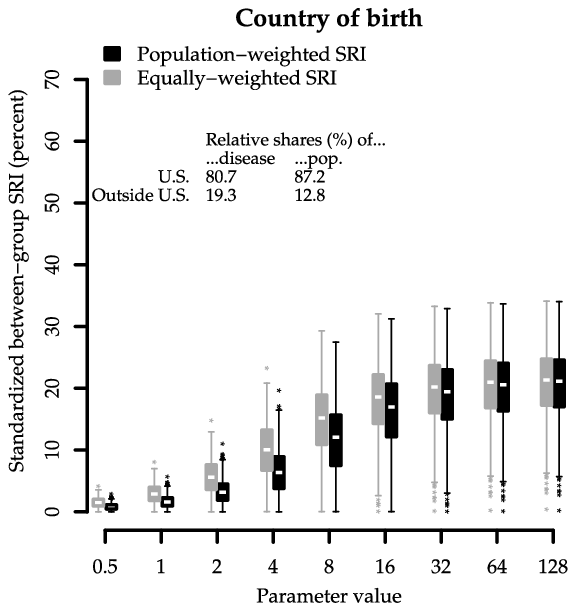}
\end{tabular}
\caption{(Continued).}
\end{figure}

%
%

As mentioned earlier, design-based standard errors obtained via Taylor
series linearization can be validated against---and are generally in
agreement with---the ones that are obtained via balanced repeated
replication and rescaled bootstrap, as shown in Table \ref{SEtab} for
the analysis by race/ethnicity.

\begin{table}
\tabcolsep=0pt
\caption{Standardized between-group SRI (in percent) for
the analysis by race/ethnicity of moderate or severe periodontitis
prevalence among U.S. adults aged 45--74, 2001--04: Comparison of
standard error (SE) for various design-based estimation methods}\label{SEtab}
\begin{tabular*}{\textwidth}{@{\extracolsep{\fill}}ld{1.3}d{1.3}d{1.3}d{2.3}d{2.3}d{2.3}d{2.3}d{2.3}d{2.3}@{}}
\hline
\multicolumn{1}{@{}l}{\textbf{Parameter value} $\bolds{(\alpha)}$} &\multicolumn{1}{c}{\textbf{0.5}} &\multicolumn{1}{c}{\textbf{1}}
&\multicolumn{1}{c}{\textbf{2}} &\multicolumn{1}{c}{\textbf{4}} &\multicolumn{1}{c}{\textbf{8}} &
\multicolumn{1}{c}{\textbf{16}} &\multicolumn{1}{c}{\textbf{32}} &\multicolumn{1}{c}{\textbf{64}} &\multicolumn{1}{c@{}}{\textbf{128}}\\
\hline
\textit{Population weighted} & & & & & & & & &\\
\quad Index &1.85 &3.67 &7.25 &13.82 &21.76 &26.56 &28.84 &29.93 &30.45\\
\quad Taylor linearization SE &0.711 &1.395 &2.662 &4.509 &5.244 &4.738
&4.349 &4.063 &3.846\\
\quad Balanced repeated&&&&&&&&&\\
\qquad replication SE &0.678 &1.330 &2.530 &4.188 &4.704
&4.388 &4.224 &4.142 &4.098\\
\quad Rescaled bootstrap SE &0.728 &1.428 &2.707 &4.393 &4.838 &4.449 &4.234
&4.137 &4.091\\[3pt]
\textit{Equally weighted} & & & & & & & & &\\
\quad Index &2.19 &4.33 &8.34 &14.87 &21.96 &26.47 &28.78 &29.90 &30.43\\
\quad Taylor linearization SE &0.822 &1.604 &3.013 &4.948 &5.783 &5.412
&5.003 &4.560 &4.060\\
\quad Balanced repeated&&&&&&&&&\\
\qquad replication SE &0.665 &1.297 &2.420 &3.888 &4.495
&4.331 &4.206 &4.137 &4.096\\
\quad Rescaled bootstrap SE &0.693 &1.352 &2.516 &3.981 &4.532 &4.355 &4.200
&4.122 &4.084
\\
\hline
\end{tabular*}
\end{table}

Notice how the two indices in Figure \ref{OHpanel} agree perfectly for
the analysis by sex, since males and females are represented almost
equally in the population. On the other hand, when the ``Other'' category
is taken into account in the analysis by race/ethnicity, the
population-weighted SRI tends to be larger than the equally-weighted
SRI for all values of the parameter $\alpha$, whereas when ``Other'' is
excluded, this ordering is reversed. This suggests that the analyst
should carefully assess the interaction between the groups' weighting
scheme and the partitioning of the population. Still, unlike in \citet
{harperetal2010}, where the effect of weighting relative to
population size versus weighting equally was examined using different
classes of indices---the MLD for the former (a GE-based HDI with the
average health outcome as the reference), but the IDisp for the latter
(a nonentropy based HDI with the least adverse health outcome as the
reference)---the SRI class of HDIs introduced in this paper provides
a unified framework for such comparative analyses, controlling more
effectively for other characteristics of the index. However, we concur
with \citet{harperetal2010} that researchers should recognize that
relying on only one HDI inevitably endorses normative judgments of one
nature or another. Though they are mostly in agreement, here, it is
clear from Figure \ref{OHpanel} that it is incumbent on researchers to
consider both the population-weighted and equally-weighted SRIs, as
well as the gradient that corresponds to increasing values of the
disparity aversion parameter.

\begin{figure}

\includegraphics{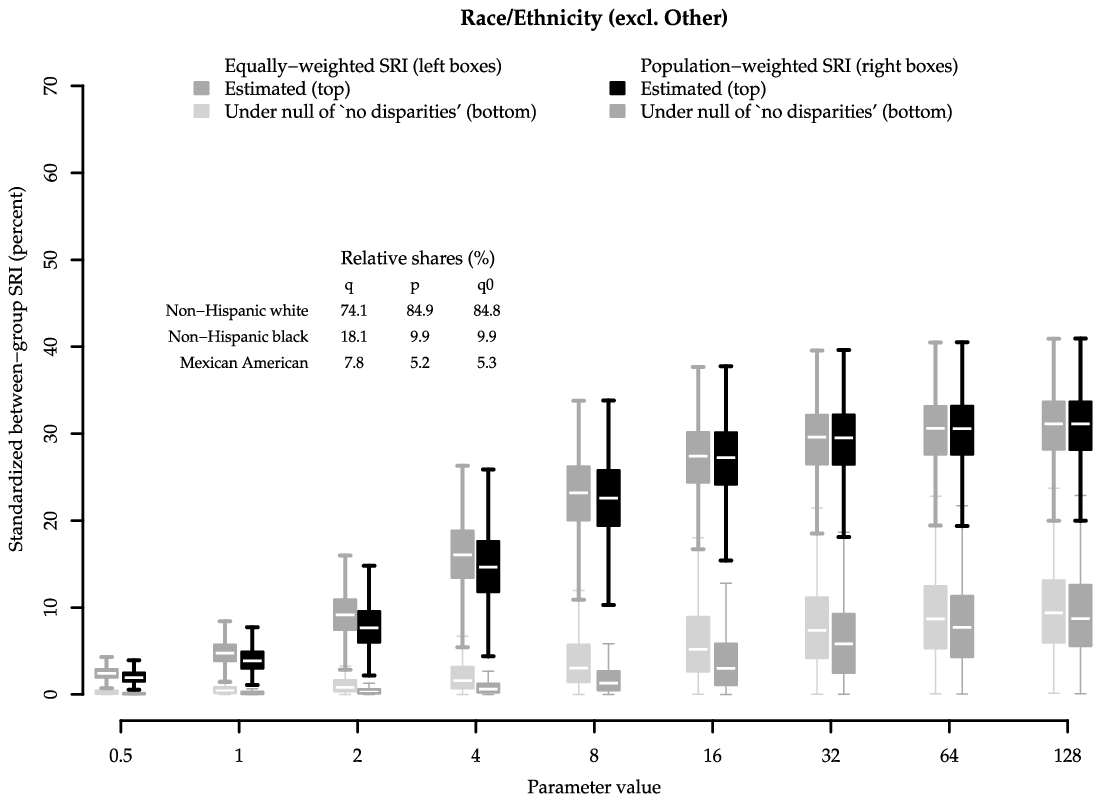}

\caption{Standardized between-group SRI for the analysis of moderate or
severe periodontitis prevalence by race/ethnicity among U.S. adults
aged 45--74, 2001--04. The population-weighted SRI uses the estimated
distributions (displayed in the offset table) for the relative shares
of population ($p_j= n_j/n$) and of disease ($q_j = y_{\cdot j}/y_{\cdot
\cdot}$) in the symmetrized R\'{e}nyi divergence $S_{\alpha}(p, q)$,
whereas the equally-weighted SRI uses $p_j = 1/m.$ The box plots are
design based, obtained via rescaled bootstrap with 500 replications.
The null hypothesis of ``no disparities'' is tested with simulated data
for which the null distribution $q_{0j}$ of disease burden is
(approximately) equal to the population shares $p_j= n_j/n$.} \label{nulltest}
\end{figure}

In Figure \ref{nulltest}, the sampling distribution of the index is
compared to one that is obtained under a null hypothesis of ``no
disparities.'' For the analysis by race/ethnicity, and without
disrupting the survey design structure, a dummy disease indicator
variable is simulated such that the relative shares of disease, $q_j =
y_{\cdot j}/y_{\cdot\cdot}$, are (approximately) equal to the given
relative population shares, $p_j= n_j/n.$ As seen in Figure \ref
{nulltest}, the resulting null and alternative distributions using the
population-weighted SRI are well separated, indicating that the null
hypothesis of ``no disparities'' would be rejected for all values of the
parameter $\alpha.$ Further, even if we were to use the
equally-weighted SRI instead (i.e., $p_j = 1/m$ instead of $p_{0j} =
n_j/n$), we would still reject the null; the overlap between the null
and alternative distributions remains minimal.

The case study in Section \ref{case-studies} illustrates how the SRI
can help examine disparities in the prevalence of moderate or severe
periodontitis among adults aged 45--74 with data from NHANES 2001--04.
This case study is relevant to HP2020 because, as stated in Section \ref
{intro}, most population-based objectives in HP2020 track a proportion
or a rate where the underlying individual-level variable has a binary
outcome, and because NHANES is the data source for approximately 1 in 7
population-based objectives in HP2020. The supplementary case study in
\citet{talih2012suppDT} provides further illustration of the proposed
methodology with continuous individual-level data on total blood
cholesterol levels among adults aged 20 and over from NHANES 2005--08.
These data track Heart Disease and Stroke objective HDS-8 in HP2020.

\textit{Caveat}. The stratified multistage probability sampling
design structure of NHANES is well documented; see \url
{http://www.cdc.gov/nchs/nhanes.htm}. While the sample weights provided
in the NHANES public-use data files reflect the unequal probabilities
of selection, they also reflect nonresponse adjustments and adjustments
to independent population controls. Therefore, strictly speaking, they
are not the true sampling weights $w_{ics}$ in (\ref{Uj}).

\section{Conclusion} \label{DISC}

In this paper we introduce a new class of HDIs, the R\'{e}nyi index
(RI), which is based on a generalized R\'{e}nyi divergence. When
standardized, the RI generalizes the Atkinson index, thus, a disparity
aversion parameter can incorporate societal values associated with
health equity. In addition, both the MLD and TI, which belong to the GE
class of HDIs, are limiting cases of the RI. Like the MLD and TI, the
RI can be symmetrized, resulting in the symmetrized R\'{e}nyi index
(SRI). We use Taylor series linearization, balanced repeated
replication, and rescaled bootstrap to examine the design-based
standard errors and bootstrapped sampling distributions for the
between-group RI and SRI in complex survey data such as NHANES. A
critical property of the RI and SRI is their invariance to the choice
of the reference used for evaluating disparities, which implies that
the index remains the same, regardless of whether we use the population
average as the reference, the group with the least adverse health
outcome, a Healthy People target, or some other reference. This
invariance property is critical to initiatives that monitor health
disparities because the identification of a reference group can be
affected by statistical reliability. An important property of the SRI
is its robustness when compared with its GE-based counterpart.

Unlike in past comparative studies, the SRI class of HDIs introduced
here provides a unified framework for ascertaining the effect of
weighting groups relative to population size versus weighting groups
equally, while controlling more effectively for other characteristics
of the index. Nonetheless, we concur with past studies that relying on
only one HDI inevitably endorses some normative judgments. Thus, it is
incumbent on the analyst who would use the SRI to consider both
population- and equally-weighted values, together with the disparity
aversion gradient. This would enable sensitivity analyses that support
development of policy recommendations that are more robust to the
numerous value judgments, both implicit and explicit, in the
measurement of health disparities. Further, although the disparity
aversion parameter $\alpha$ in the standardized SRI is treated in this
paper as a ``tuning'' parameter, future work could, instead, determine
the parameter $\alpha$ from global variables such as cost of treatment,
availability of health care resources, and other structural factors
discussed in \citet{fleurbaeyschokkaert2009}.

\section*{Acknowledgments}
The author thanks Luisa Borrell (CUNY)
and Elizabeth Jackson (NCHS) for a discussion of the periodontitis case
definitions in HP2020 objective OH-5, and Kimberly Rosendorf (NCHS) for
input on findings on total blood cholesterol levels (HP2020 objective
HDS-8). Jeff Pearcy (NCHS) and Yukiko Asada (Dalhousie University,
Canada) discussed the interpretation of the parameter $\alpha$ as a
disparity aversion parameter for the standardized SRI. Comments
received from Van Parsons (NCHS) helped improve the section on variance
estimation. The support and input received from Rebecca Hines, Chief of
the Health Promotion Statistics Branch at NCHS, are gratefully
acknowledged. Comments from Richard Klein, Jennifer Madans (NCHS), the
journal editors, and anonymous reviewers have vastly improved the
presentation of the material in the paper.

\begin{supplement}[id=suppA]
\sname{Supplement A}
\stitle{Technical appendix: Decomposability\\}
\slink[doi]{10.1214/12-AOAS621SUPPA} 
\sdatatype{.pdf}
\sfilename{aoas621\_suppa.pdf}
\sdescription{Expressions and variance calculations for the total or
aggregate RI and SRI and their within-group components when
individual-level data are continuous.}
\end{supplement}

\begin{supplement}[id=suppA]
\sname{Supplement B}
\stitle{Additional case study from NHANES\\}
\slink[doi]{10.1214/12-AOAS621SUPPB} 
\sdatatype{.pdf}
\sfilename{aoas621\_suppb.pdf}
\sdescription{Disparities in mean total blood cholesterol levels ($\mu
$g/dL) in U.S. adults aged 20 and over, 2005--08.}
\end{supplement}

\begin{supplement}[id=suppA]
\sname{Supplement C}
\stitle{R syntax and output files\\}
\slink[doi,text={10.1214/12-\break AOAS621SUPPC}]{10.1214/12-AOAS621SUPPC} 
\sdatatype{.zip}
\sfilename{aoas621\_suppc.zip}
\sdescription{Syntax and output from case studies comparing the
equally-weighted and population-weighted RI and SRI; their
group-specific, between-, and within-group components; and their
design-based standard errors and sampling distributions, obtained via
Taylor series linearization, balanced repeated replication, and
rescaled bootstrap. Syntax is reverse-compatible with that in
\citeauthor{borrelltalih2011} (\citeyear{borrelltalih2011,borrelltalih2012}).}
\end{supplement}


%

\printaddresses


\begin{thebibliography}{53}

\bibitem[\protect\citeauthoryear{Ali and Silvey}{1966}]{alisilvey1966}
\begin{barticle}[mr]
\bauthor{\bsnm{Ali},~\bfnm{S.~M.}\binits{S.~M.}} \AND
\bauthor{\bsnm{Silvey},~\bfnm{S.~D.}\binits{S.~D.}}
(\byear{1966}).
\btitle{A general class of coefficients of divergence of one distribution from
another}.
\bjournal{J. Roy. Statist. Soc. Ser. B}
\bvolume{28}
\bpages{131--142}.
\bid{issn={0035-9246}, mr={0196777}}
\bptok{imsref}%
\end{barticle}
\endbibitem

\bibitem[\protect\citeauthoryear{Atkinson}{1970}]{atkinson1970}
\begin{barticle}[mr]
\bauthor{\bsnm{Atkinson},~\bfnm{Anthony~B.}\binits{A.~B.}}
(\byear{1970}).
\btitle{On the measurement of inequality}.
\bjournal{J. Econom. Theory}
\bvolume{2}
\bpages{244--263}.
\bid{issn={0022-0531}, mr={0449508}}
\bptok{imsref}%
\end{barticle}
\endbibitem

\bibitem[\protect\citeauthoryear{Biewen and
Jenkins}{2006}]{biewenjenkins2006}
\begin{barticle}[author]
\bauthor{\bsnm{Biewen},~\bfnm{Martin}\binits{M.}} \AND
\bauthor{\bsnm{Jenkins},~\bfnm{Stephen~P.}\binits{S.~P.}}
(\byear{2006}).
\btitle{Variance estimation for generalized entropy and {Atkinson} inequality
indices: The complex survey data case}.
\bjournal{Oxford Bulletin of Economics and Statistics}
\bvolume{68}
\bpages{371--383}.
\bptok{imsref}%
\end{barticle}
\endbibitem

\bibitem[\protect\citeauthoryear{Borrell and Talih}{2012}]{borrelltalih2012}
\begin{barticle}[pbm]
\bauthor{\bsnm{Borrell},~\bfnm{Luisa~N.}\binits{L.~N.}} \AND
\bauthor{\bsnm{Talih},~\bfnm{Makram}\binits{M.}}
(\byear{2012}).
\btitle{Examining periodontal disease disparities among U.S. adults 20 years of
age and older: NHANES III (1988--1994) and NHANES 1999--2004}.
\bjournal{Public Health Rep.}
\bvolume{127}
\bpages{497--506}.
\bid{issn={1468-2877}, pmcid={3407849}, pmid={22942467}}
\bptok{imsref}%
\end{barticle}
\endbibitem

\bibitem[\protect\citeauthoryear{Borrell and Talih}{2011}]{borrelltalih2011}
\begin{barticle}[mr]
\bauthor{\bsnm{Borrell},~\bfnm{Luisa~N.}\binits{L.~N.}} \AND
\bauthor{\bsnm{Talih},~\bfnm{Makram}\binits{M.}}
(\byear{2011}).
\btitle{A symmetrized {T}heil index measure of health disparities: An example
using dental caries in {U}.{S}. children and adolescents}.
\bjournal{Stat. Med.}
\bvolume{30}
\bpages{277--290}.
\bid{doi={10.1002/sim.4114}, issn={0277-6715}, mr={2758878}}
\bptok{imsref}%
\end{barticle}
\endbibitem

\bibitem[\protect\citeauthoryear{Bourguignon}{1979}]{bourguignon1979}
\begin{barticle}[mr]
\bauthor{\bsnm{Bourguignon},~\bfnm{Fran{\c{c}}ois}\binits{F.}}
(\byear{1979}).
\btitle{Decomposable income inequality measures}.
\bjournal{Econometrica}
\bvolume{47}
\bpages{901--920}.
\bid{doi={10.2307/1914138}, issn={0012-9682}, mr={0537636}}
\bptok{imsref}%
\end{barticle}
\endbibitem

\bibitem[\protect\citeauthoryear{Braveman}{2006}]{braveman2006}
\begin{barticle}[pbm]
\bauthor{\bsnm{Braveman},~\bfnm{Paula}\binits{P.}}
(\byear{2006}).
\btitle{Health disparities and health equity: Concepts and measurement}.
\bjournal{Annu. Rev. Public Health}
\bvolume{27}
\bpages{167--194}.
\bid{doi={10.1146/annurev.publhealth.27.021405.102103}, issn={0163-7525},
pmid={16533114}}
\bptok{imsref}%
\end{barticle}
\endbibitem

\bibitem[\protect\citeauthoryear{Cheng, Han and Gansky}{2008}]{chengetal2008}
\begin{barticle}[pbm]
\bauthor{\bsnm{Cheng},~\bfnm{Nancy~F.}\binits{N.~F.}},
\bauthor{\bsnm{Han},~\bfnm{Pamela~Z.}\binits{P.~Z.}} \AND
\bauthor{\bsnm{Gansky},~\bfnm{Stuart~A.}\binits{S.~A.}}
(\byear{2008}).
\btitle{Methods and software for estimating health disparities: The case of
children's oral health}.
\bjournal{Am. J. Epidemiol.}
\bvolume{168}
\bpages{906--914}.
\bid{doi={10.1093/aje/kwn207}, issn={1476-6256}, mid={NIHMS63747},
pii={kwn207}, pmcid={2597673}, pmid={18779387}}
\bptok{imsref}%
\end{barticle}
\endbibitem

\bibitem[\protect\citeauthoryear{Chernoff}{1952}]{chernoff1952}
\begin{barticle}[mr]
\bauthor{\bsnm{Chernoff},~\bfnm{Herman}\binits{H.}}
(\byear{1952}).
\btitle{A measure of asymptotic efficiency for tests of a hypothesis based on
the sum of observations}.
\bjournal{Ann. Math. Statistics}
\bvolume{23}
\bpages{493--507}.
\bid{issn={0003-4851}, mr={0057518}}
\bptok{imsref}%
\end{barticle}
\endbibitem

\bibitem[\protect\citeauthoryear{Cichocki and
Amari}{2010}]{cichockiamari2010}
\begin{barticle}[mr]
\bauthor{\bsnm{Cichocki},~\bfnm{Andrzej}\binits{A.}} \AND
\bauthor{\bsnm{Amari},~\bfnm{Shun-ichi}\binits{S.-i.}}
(\byear{2010}).
\btitle{Families of alpha- beta- and gamma-divergences: Flexible and robust
measures of similarities}.
\bjournal{Entropy}
\bvolume{12}
\bpages{1532--1568}.
\bid{doi={10.3390/e12061532}, issn={1099-4300}, mr={2659408}}
\bptok{imsref}%
\end{barticle}
\endbibitem

\bibitem[\protect\citeauthoryear{Cowell, Davidson and
Flachaire}{2011}]{cowelletal2011}
\begin{bmisc}[author]
\bauthor{\bsnm{Cowell},~\bfnm{Frank~A.}\binits{F.~A.}},
\bauthor{\bsnm{Davidson},~\bfnm{Russel}\binits{R.}} \AND
\bauthor{\bsnm{Flachaire},~\bfnm{Emmanuel}\binits{E.}}
(\byear{2011}).
\bhowpublished{Goodness of fit: An axiomatic approach.
Groupement de Recherche en Economie Quantitative D'Aix-Marseille
(GREQAM) DT 2011-50. Available at
\texttt{\href{http://halshs.archives-ouvertes.fr/docs/00/63/90/75/PDF/DTGREQAM2011\_50.pdf}{http://halshs.archives-ouvertes.fr/}
\href{http://halshs.archives-ouvertes.fr/docs/00/63/90/75/PDF/DTGREQAM2011\_50.pdf}{docs/00/63/90/75/PDF/DTGREQAM2011\_50.pdf}}.}
\bptok{imsref}%
\end{bmisc}
\endbibitem

\bibitem[\protect\citeauthoryear{Cowell and Kuga}{1981}]{cowellkuga1980}
\begin{barticle}[mr]
\bauthor{\bsnm{Cowell},~\bfnm{Frank~A.}\binits{F.~A.}} \AND
\bauthor{\bsnm{Kuga},~\bfnm{Kiyoshi}\binits{K.}}
(\byear{1981}).
\btitle{Additivity and the entropy concept: An axiomatic approach to inequality
measurement}.
\bjournal{J. Econom. Theory}
\bvolume{25}
\bpages{131--143}.
\bid{doi={10.1016/0022-0531(81)90020-X}, issn={0022-0531}, mr={0636017}}
\bptnote{check year}%
\bptok{imsref}%
\end{barticle}
\endbibitem

\bibitem[\protect\citeauthoryear{Cressie and Read}{1984}]{cressieread1984}
\begin{barticle}[mr]
\bauthor{\bsnm{Cressie},~\bfnm{Noel}\binits{N.}} \AND
\bauthor{\bsnm{Read},~\bfnm{Timothy R.~C.}\binits{T.~R.~C.}}
(\byear{1984}).
\btitle{Multinomial goodness-of-fit tests}.
\bjournal{J. Roy. Statist. Soc. Ser.~B}
\bvolume{46}
\bpages{440--464}.
\bid{issn={0035-9246}, mr={0790631}}
\bptok{imsref}%
\end{barticle}
\endbibitem

\bibitem[\protect\citeauthoryear{Elbers et~al.}{2008}]{elbersetal2008}
\begin{barticle}[author]
\bauthor{\bsnm{Elbers},~\bfnm{Chris}\binits{C.}},
\bauthor{\bsnm{Lanjouw},~\bfnm{Peter}\binits{P.}},
\bauthor{\bsnm{Mistiaen},~\bfnm{Johan~A.}\binits{J.~A.}} \AND
\bauthor{\bsnm{{\"{O}}zler},~\bfnm{Berk}\binits{B.}}
(\byear{2008}).
\btitle{Reinterpreting between-group inequality}.
\bjournal{Journal of Economic Inequality}
\bvolume{6}
\bpages{231--245}.
\bptok{imsref}%
\end{barticle}
\endbibitem

\bibitem[\protect\citeauthoryear{Fay}{1989}]{fay1989}
\begin{binproceedings}[author]
\bauthor{\bsnm{Fay},~\bfnm{R.~E.}\binits{R.~E.}}
(\byear{1989}).
\btitle{Theoretical application of weighting for variance calculation}.
In \bbooktitle{Proceedings of the Section on Survey Research Methods}
\bpages{212--217}.
\bpublisher{Amer. Statist. Assoc.}, \baddress{ Alexandria, VA}.
\bptok{imsref}%
\end{binproceedings}
\endbibitem

\bibitem[\protect\citeauthoryear{Firebaugh}{1999}]{firebaugh1999}
\begin{barticle}[author]
\bauthor{\bsnm{Firebaugh},~\bfnm{Glenn}\binits{G.}}
(\byear{1999}).
\btitle{Empirics of world income inequality}.
\bjournal{American Journal of Sociology}
\bvolume{104}
\bpages{1597--1630}.
\bptok{imsref}%
\end{barticle}
\endbibitem

\bibitem[\protect\citeauthoryear{Fleurbaey and
Schokkaert}{2009}]{fleurbaeyschokkaert2009}
\begin{barticle}[pbm]
\bauthor{\bsnm{Fleurbaey},~\bfnm{Marc}\binits{M.}} \AND
\bauthor{\bsnm{Schokkaert},~\bfnm{Erik}\binits{E.}}
(\byear{2009}).
\btitle{Unfair inequalities in health and health care}.
\bjournal{J.~Health Econ.}
\bvolume{28}
\bpages{73--90}.
\bid{doi={10.1016/j.jhealeco.2008.07.016}, issn={0167-6296},
pii={S0167-6296(08)00102-1}, pmid={18829124}}
\bptok{imsref}%
\end{barticle}
\endbibitem



\bibitem[\protect\citeauthoryear{Frohlich and
Potvin}{2008}]{frohlichpotvin2008}
\begin{barticle}[pbm]
\bauthor{\bsnm{Frohlich},~\bfnm{Katherine~L.}\binits{K.~L.}} \AND
\bauthor{\bsnm{Potvin},~\bfnm{Louise}\binits{L.}}
(\byear{2008}).
\btitle{Transcending the known in public health practice: The inequality
paradox: The population approach and vulnerable populations}.
\bjournal{Am. J. Public Health}
\bvolume{98}
\bpages{216--221}.
\bid{doi={10.2105/AJPH.2007.114777}, issn={1541-0048}, pii={AJPH.2007.114777},
pmcid={2376882}, pmid={18172133}}
\bptok{imsref}%
\end{barticle}
\endbibitem

\bibitem[\protect\citeauthoryear{Fujisawa and
Eguchi}{2008}]{fujisawaeguchi2008}
\begin{barticle}[mr]
\bauthor{\bsnm{Fujisawa},~\bfnm{Hironori}\binits{H.}} \AND
\bauthor{\bsnm{Eguchi},~\bfnm{Shinto}\binits{S.}}
(\byear{2008}).
\btitle{Robust parameter estimation with a small bias against heavy
contamination}.
\bjournal{J. Multivariate Anal.}
\bvolume{99}
\bpages{2053--2081}.
\bid{doi={10.1016/j.jmva.2008.02.004}, issn={0047-259X}, mr={2466551}}
\bptok{imsref}%
\end{barticle}
\endbibitem

\bibitem[\protect\citeauthoryear{Green and
Fielding}{2011}]{greenfielding2011}
\begin{barticle}[pbm]
\bauthor{\bsnm{Green},~\bfnm{Lawrence~W.}\binits{L.~W.}} \AND
\bauthor{\bsnm{Fielding},~\bfnm{Jonathan}\binits{J.}}
(\byear{2011}).
\btitle{The U.S. healthy people initiative: Its genesis and its
sustainability}.
\bjournal{Annu. Rev. Public Health}
\bvolume{32}
\bpages{451--470}.
\bid{doi={10.1146/annurev-publhealth-031210-101148}, issn={1545-2093},
pmid={21417753}}
\bptok{imsref}%
\end{barticle}
\endbibitem

\bibitem[\protect\citeauthoryear{Harper et~al.}{2008}]{harperetal2008}
\begin{barticle}[pbm]
\bauthor{\bsnm{Harper},~\bfnm{Sam}\binits{S.}},
\bauthor{\bsnm{Lynch},~\bfnm{John}\binits{J.}},
\bauthor{\bsnm{Meersman},~\bfnm{Stephen~C.}\binits{S.~C.}},
\bauthor{\bsnm{Breen},~\bfnm{Nancy}\binits{N.}},
\bauthor{\bsnm{Davis},~\bfnm{William~W.}\binits{W.~W.}} \AND
\bauthor{\bsnm{Reichman},~\bfnm{Marsha~E.}\binits{M.~E.}}
(\byear{2008}).
\btitle{An overview of methods for monitoring social disparities in cancer with
an example using trends in lung cancer incidence by area-socioeconomic
position and race-ethnicity, 1992--2004}.
\bjournal{Am. J. Epidemiol.}
\bvolume{167}
\bpages{889--899}.
\bid{doi={10.1093/aje/kwn016}, issn={1476-6256}, mid={NIHMS48152},
pii={kwn016}, pmcid={2409988}, pmid={18344513}}
\bptok{imsref}%
\end{barticle}
\endbibitem

\bibitem[\protect\citeauthoryear{Harper et~al.}{2010}]{harperetal2010}
\begin{barticle}[author]
\bauthor{\bsnm{Harper},~\bfnm{Sam}\binits{S.}},
\bauthor{\bsnm{King},~\bfnm{Nicholas~B.}\binits{N.~B.}},
\bauthor{\bsnm{Meersman},~\bfnm{Stephen~C.}\binits{S.~C.}},
\bauthor{\bsnm{Reichman},~\bfnm{Marsha~E.}\binits{M.~E.}},
\bauthor{\bsnm{Breen},~\bfnm{Nancy}\binits{N.}} \AND
\bauthor{\bsnm{Lynch},~\bfnm{John}\binits{J.}}
(\byear{2010}).
\btitle{Implicit value judgments in the measurement of health inequalities}.
\bjournal{Milbank Quaterly}
\bvolume{88}
\bpages{4--29}.
\bptok{imsref}%
\end{barticle}
\endbibitem

\bibitem[\protect\citeauthoryear{Haughton and Khander}{2009}]{worldbank2009}
\begin{bbook}[author]
\bauthor{\bsnm{Haughton},~\bfnm{Jonathan}\binits{J.}} \AND
\bauthor{\bsnm{Khander},~\bfnm{Shahidur~R.}\binits{S.~R.}}
(\byear{2009}).
\btitle{Handbook on Poverty and Inequality}.
\bpublisher{The World Bank}, \blocation{Washington, DC}.
\bptok{imsref}%
\end{bbook}
\endbibitem

\bibitem[\protect\citeauthoryear{Judkins}{1990}]{judkins1990}
\begin{barticle}[author]
\bauthor{\bsnm{Judkins},~\bfnm{D.~R.}\binits{D.~R.}}
(\byear{1990}).
\btitle{Fay's method for variance estimation}.
\bjournal{Journal of Official Statistics}
\bvolume{6}
\bpages{223--239}.
\bptok{imsref}%
\end{barticle}
\endbibitem

\bibitem[\protect\citeauthoryear{Keppel, Pearcy and
Klein}{2004}]{keppelpearcyklein2004}
\begin{barticle}[pbm]
\bauthor{\bsnm{Keppel},~\bfnm{Kenneth~G.}\binits{K.~G.}},
\bauthor{\bsnm{Pearcy},~\bfnm{Jeffrey~N.}\binits{J.~N.}} \AND
\bauthor{\bsnm{Klein},~\bfnm{Richard~J.}\binits{R.~J.}}
(\byear{2004}).
\btitle{Measuring progress in Healthy People 2010}.
\bjournal{Healthy People 2010 Stat. Notes}
\bvolume{25}
\bpages{1--16}.
\bid{pmid={15446274}}
\bptok{imsref}%
\end{barticle}
\endbibitem

\bibitem[\protect\citeauthoryear{Keppel
et~al.}{2005}]{keppelpamuklynchetal2005}
\begin{bbook}[author]
\bauthor{\bsnm{Keppel},~\bfnm{Kenneth}\binits{K.}},
\bauthor{\bsnm{Pamuk},~\bfnm{Elsie}\binits{E.}},
\bauthor{\bsnm{Lynch},~\bfnm{John}\binits{J.}},
\bauthor{\bsnm{{Carter-Pokras}},~\bfnm{Olivia}\binits{O.}},
\bauthor{\bsnm{Kim},~\bfnm{Insun}\binits{I.}},
\bauthor{\bsnm{Mays},~\bfnm{Vickie}\binits{V.}},
\bauthor{\bsnm{Pearcy},~\bfnm{Jeffrey}\binits{J.}},
\bauthor{\bsnm{Schoenbach},~\bfnm{Victor}\binits{V.}} \AND
\bauthor{\bsnm{Weissman},~\bfnm{Joel~S.}\binits{J.~S.}}
(\byear{2005}).
\btitle{Methodological Issues in Measuring Health Disparities}.
\bseries{Vital and Health Statistics, Series 2}
\bvolume{141}.
\bpublisher{National Center for Health Statistics}, \baddress{Hyattsville, MD}.
\bptok{imsref}%
\end{bbook}
\endbibitem

\bibitem[\protect\citeauthoryear{Kullback and
Leibler}{1951}]{kullbackleibler1951}
\begin{barticle}[mr]
\bauthor{\bsnm{Kullback},~\bfnm{S.}\binits{S.}} \AND
\bauthor{\bsnm{Leibler},~\bfnm{R.~A.}\binits{R.~A.}}
(\byear{1951}).
\btitle{On information and sufficiency}.
\bjournal{Ann. Math. Statistics}
\bvolume{22}
\bpages{79--86}.
\bid{issn={0003-4851}, mr={0039968}}
\bptok{imsref}%
\end{barticle}
\endbibitem

\bibitem[\protect\citeauthoryear{Levy, Chemerynski and
Tuchmann}{2006}]{levyetal2006}
\begin{bmisc}[author]
\bauthor{\bsnm{Levy},~\bfnm{Jonathan~I.}\binits{J.~I.}},
\bauthor{\bsnm{Chemerynski},~\bfnm{Susan~M.}\binits{S.~M.}} \AND
\bauthor{\bsnm{Tuchmann},~\bfnm{Jessica~L.}\binits{J.~L.}}
(\byear{2006}).
\bhowpublished{Incorporating concepts of inequality and inequity into health benefits
analysis. \textit{International Journal of Equity in Health} \textbf{5}. Available at DOI:\doiurl{10.1186/1475-9276-5-2}.}
\bptok{imsref}%
\end{bmisc}
\endbibitem

\bibitem[\protect\citeauthoryear{Lumley}{2004}]{lumley2004}
\begin{barticle}[author]
\bauthor{\bsnm{Lumley},~\bfnm{T.}\binits{T.}}
(\byear{2004}).
\btitle{Analysis of complex survey samples}.
\bjournal{Journal of Statistical Software}
\bvolume{9}
\bpages{1--19}.
\bptok{imsref}%
\end{barticle}
\endbibitem

\bibitem[\protect\citeauthoryear{Lumley}{2011}]{Rsurvey2011}
\begin{bmisc}[author]
\bauthor{\bsnm{Lumley},~\bfnm{T.}\binits{T.}}
(\byear{2011}).
\bhowpublished{``Survey'': Analysis of complex survey samples.
R package version 3.26.}
\bptok{imsref}%
\end{bmisc}
\endbibitem

\bibitem[\protect\citeauthoryear{Mackenbach and
Kunst}{1997}]{mackenbachkunst1997}
\begin{barticle}[pbm]
\bauthor{\bsnm{Mackenbach},~\bfnm{J.~P.}\binits{J.~P.}} \AND
\bauthor{\bsnm{Kunst},~\bfnm{A.~E.}\binits{A.~E.}}
(\byear{1997}).
\btitle{Measuring the magnitude of socio-economic inequalities in health: An
overview of available measures illustrated with two examples from Europe}.
\bjournal{Soc. Sci. Med.}
\bvolume{44}
\bpages{757--771}.
\bid{issn={0277-9536}, pii={S0277953696000731}, pmid={9080560}}
\bptok{imsref}%
\end{barticle}
\endbibitem

\bibitem[\protect\citeauthoryear{Magdalou and Nock}{2011}]{magdalounock2011}
\begin{barticle}[mr]
\bauthor{\bsnm{Magdalou},~\bfnm{Brice}\binits{B.}} \AND
\bauthor{\bsnm{Nock},~\bfnm{Richard}\binits{R.}}
(\byear{2011}).
\btitle{Income distributions and decomposable divergence measures}.
\bjournal{J. Econom. Theory}
\bvolume{146}
\bpages{2440--2454}.
\bid{doi={10.1016/j.jet.2011.06.017}, issn={0022-0531}, mr={2887794}}
\bptok{imsref}%
\end{barticle}
\endbibitem

\bibitem[\protect\citeauthoryear{Mart{\'{\i}}nez-Camblor}{2007}]{martinez-camblor2007}
\begin{barticle}[mr]
\bauthor{\bsnm{Mart{\'{\i}}nez-Camblor},~\bfnm{Pablo}\binits{P.}}
(\byear{2007}).
\btitle{Central limit theorems for {$S$}-{G}ini and {T}heil inequality
coefficients}.
\bjournal{Rev. Colombiana Estad\'\i st.}
\bvolume{30}
\bpages{287--300}.
\bid{issn={0120-1751}, mr={2422860}}
\bptok{imsref}%
\end{barticle}
\endbibitem

\bibitem[\protect\citeauthoryear{Mc{C}arthy}{1969}]{mccarthy1969}
\begin{barticle}[author]
\bauthor{\bsnm{Mc{C}arthy},~\bfnm{P.~J.}\binits{P.~J.}}
(\byear{1969}).
\btitle{Pseudo-replication: Half samples}.
\bjournal{{Revue de l'Institut International de Statistique---Review of the
International Statistical Institute}}
\bvolume{37}
\bpages{239--264}.
\bptok{imsref}%
\end{barticle}
\endbibitem


\bibitem[\protect\citeauthoryear{{National Center for Health
Statistics}}{2011}]{fr2011}
\begin{bmisc}[author]
\borganization{National Center for Health Statistics}.
(\byear{2011}).
\bhowpublished{Healthy People 2010 Final Review.
National Center for Health Statistics, Hyattsville,
MD.}
\bptok{imsref}%
\end{bmisc}
\endbibitem


\bibitem[\protect\citeauthoryear{Page and Eke}{2007}]{pageeke2007}
\begin{barticle}[pbm]
\bauthor{\bsnm{Page},~\bfnm{Roy~C.}\binits{R.~C.}} \AND
\bauthor{\bsnm{Eke},~\bfnm{Paul~I.}\binits{P.~I.}}
(\byear{2007}).
\btitle{Case definitions for use in population-based surveillance of
periodontitis}.
\bjournal{J. Periodontol.}
\bvolume{78}
\bpages{1387--1399}.
\bid{doi={10.1902/jop.2007.060264}, issn={0022-3492}, pmid={17608611}}
\bptok{imsref}%
\end{barticle}
\endbibitem

\bibitem[\protect\citeauthoryear{Pearcy and Keppel}{2002}]{pearcykeppel2002}
\begin{barticle}[author]
\bauthor{\bsnm{Pearcy},~\bfnm{Jeff~N.}\binits{J.~N.}} \AND
\bauthor{\bsnm{Keppel},~\bfnm{Ken~G.}\binits{K.~G.}}
(\byear{2002}).
\btitle{A summary measure of health disparity}.
\bjournal{Public Health Reports}
\bvolume{117}
\bpages{273--280}.
\bptok{imsref}%
\end{barticle}
\endbibitem

\bibitem[\protect\citeauthoryear{Pollard}{2002}]{pollard2002}
\begin{bbook}[author]
\bauthor{\bsnm{Pollard},~\bfnm{David~E.}\binits{D.~E.}}
(\byear{2002}).
\btitle{A User's Guide to Measure Theoretic Probability}.
\bpublisher{Cambridge Univ. Press}, \blocation{Cambridge, UK}.
\bptok{imsref}%
\end{bbook}
\endbibitem

\bibitem[\protect\citeauthoryear{{R Development Core Team}}{2011}]{R2011}
\begin{bmisc}[author]
\borganization{R Development Core Team}.
(\byear{2011}).
\bhowpublished{\textit{R: A Language and Environment for Statistical Computing}.
R Foundation for Statistical Computing, Vienna,
Austria. {ISBN} 3-900051-07-0. Available at \url{http://www.R-project.org}.}
\bptok{imsref}%
\end{bmisc}
\endbibitem


\bibitem[\protect\citeauthoryear{Rao and Wu}{1988}]{raowu1988}
\begin{barticle}[mr]
\bauthor{\bsnm{Rao},~\bfnm{J.~N.~K.}\binits{J.~N.~K.}} \AND
\bauthor{\bsnm{Wu},~\bfnm{C.~F.~J.}\binits{C.~F.~J.}}
(\byear{1988}).
\btitle{Resampling inference with complex survey data}.
\bjournal{J. Amer. Statist. Assoc.}
\bvolume{83}
\bpages{231--241}.
\bid{issn={0162-1459}, mr={0941020}}
\bptok{imsref}%
\end{barticle}
\endbibitem

\bibitem[\protect\citeauthoryear{Rao, Wu and Yue}{1992}]{raoetal1992}
\begin{barticle}[author]
\bauthor{\bsnm{Rao},~\bfnm{J.~N.~K.}\binits{J.~N.~K.}},
\bauthor{\bsnm{Wu},~\bfnm{C.~F.~J.}\binits{C.~F.~J.}} \AND
\bauthor{\bsnm{Yue},~\bfnm{K.}\binits{K.}}
(\byear{1992}).
\btitle{Some recent work in resampling methods}.
\bjournal{Survey Methodology}
\bvolume{18}
\bpages{209--217}.
\bptok{imsref}%
\end{barticle}
\endbibitem

\bibitem[\protect\citeauthoryear{R{\'{e}}nyi}{1960}]{renyi1960}
\begin{binproceedings}[author]
\bauthor{\bsnm{R{\'{e}}nyi},~\bfnm{Alfr{\'{e}}d}\binits{A.}}
(\byear{1960}).
\btitle{On measures of entropy and information}.
In \bbooktitle{Proc. 4th Berkeley Sympos. Math. Statist. and Prob.}
\bpages{547--561}.
\bpublisher{Univ. California Press}, \baddress{Berkeley, CA}.
\bid{mr={0132570}}
\bptok{imsref}%
\end{binproceedings}
\endbibitem

\bibitem[\protect\citeauthoryear{Rose}{1985}]{rose1985}
\begin{barticle}[author]
\bauthor{\bsnm{Rose},~\bfnm{Geoffrey}\binits{G.}}
(\byear{1985}).
\btitle{Sick individuals and sick populations}.
\bjournal{International Journal of Epidemiology}
\bvolume{14}
\bpages{32--38}.
\bptok{imsref}%
\end{barticle}
\endbibitem

\bibitem[\protect\citeauthoryear{Shorrocks}{1980}]{shorrocks1980}
\begin{barticle}[mr]
\bauthor{\bsnm{Shorrocks},~\bfnm{A.~F.}\binits{A.~F.}}
(\byear{1980}).
\btitle{The class of additively decomposable inequality measures}.
\bjournal{Econometrica}
\bvolume{48}
\bpages{613--625}.
\bid{doi={10.2307/1913126}, issn={0012-9682}, mr={0573315}}
\bptok{imsref}%
\end{barticle}
\endbibitem

\bibitem[\protect\citeauthoryear{Talih}{2013a}]{talih2012suppTA}
\begin{bmisc}[author]
\bauthor{\bsnm{Talih},~\bfnm{Makram}\binits{M.}}
(\byear{2013}a).
\bhowpublished{Supplement to ``A reference-invariant health disparity index based on
R\'{e}nyi divergence---technical appendix.'' DOI:\doiurl{10.1214/12-AOAS621SUPPA}.}
\bptok{imsref}%
\end{bmisc}
\endbibitem

\bibitem[\protect\citeauthoryear{Talih}{2013b}]{talih2012suppDT}
\begin{bmisc}[author]
\bauthor{\bsnm{Talih},~\bfnm{Makram}\binits{M.}}
(\byear{2013}b).
\bhowpublished{Supplement to ``A reference-invariant health disparity index based on
R\'{e}nyi divergence---additional case study from NHANES.''
DOI:\doiurl{10.1214/12-AOAS621SUPPB}.}
\bptok{imsref}%
\end{bmisc}
\endbibitem

\bibitem[\protect\citeauthoryear{Talih}{2013c}]{talih2012suppRR}
\begin{bmisc}[author]
\bauthor{\bsnm{Talih},~\bfnm{Makram}\binits{M.}}
(\byear{2013}c).
\bhowpublished{Supplement to ``A reference-invariant health disparity index based on
R\'{e}nyi divergence---R syntax and output files.''
DOI:\doiurl{10.1214/12-AOAS621SUPPC}.}
\bptok{imsref}%
\end{bmisc}
\endbibitem




\bibitem[\protect\citeauthoryear{Theil}{1967}]{theil1967}
\begin{bbook}[author]
\bauthor{\bsnm{Theil},~\bfnm{Henri}\binits{H.}}
(\byear{1967}).
\btitle{Economics and Information Theory}.
\bpublisher{North Holland}, \blocation{Amsterdam, Netherlands}.
\bptok{imsref}%
\end{bbook}
\endbibitem

\bibitem[\protect\citeauthoryear{{U.S. Department of Health and Human
Services}}{2000}]{hp20102000}
\begin{bmisc}[author]
\borganization{U.S. Department of Health and Human Services}.
(\byear{2000}).
\bhowpublished{Healthy People 2010, 2nd ed: With Understanding and Improving Health
and Objectives for Improving Health,  Vol.~2.
U.S. Government Printing Office, Washington, DC}.
\bptok{imsref}%
\end{bmisc}
\endbibitem

\bibitem[\protect\citeauthoryear{{U.S. Department of Health and Human
Services}}{2006}]{mcr2006}
\begin{bmisc}[author]
\borganization{U.S. Department of Health and Human Services}.
(\byear{2006}).
\bhowpublished{Healthy People 2010 Midcourse Review.
U.S. Government Printing Office, Washington, DC}.
\bptok{imsref}%
\end{bmisc}
\endbibitem





\bibitem[\protect\citeauthoryear{{van Erven}}{2010}]{vanErven2010}
\begin{bmisc}[author]
\bauthor{\bsnm{{van Erven}},~\bfnm{Tim A.~L.}\binits{T.~A.~L.}}
(\byear{2010}).
\bhowpublished{When data compression and statistics disagree: Two frequentist
challenges for the minimum description length principle. Ph.D. thesis, Leiden University---CWI, the Netherlands. {ISBN} 978-90-9025673-3.
Available at \url{http://hdl.handle.net/1887/15879}.}
\bptok{imsref}%
\end{bmisc}
\endbibitem

\bibitem[\protect\citeauthoryear{Wagstaff, Paci and van
Doorslaer}{1991}]{wagstaffetal1991}
\begin{barticle}[pbm]
\bauthor{\bsnm{Wagstaff},~\bfnm{A.}\binits{A.}},
\bauthor{\bsnm{Paci},~\bfnm{P.}\binits{P.}} \AND \bauthor{\bparticle{van}
\bsnm{Doorslaer},~\bfnm{E.}\binits{E.}}
(\byear{1991}).
\btitle{On the measurement of inequalities in health}.
\bjournal{Soc. Sci. Med.}
\bvolume{33}
\bpages{545--557}.
\bid{issn={0277-9536}, pmid={1962226}}
\bptok{imsref}%
\end{barticle}
\endbibitem

\end{thebibliography}
\end{document}